\documentclass[sn-basic,iicol,10pt]{sn-jnl}% Default with double column layout

\jyear{2022}%

\theoremstyle{thmstyleone}%
\theoremstyle{thmstyletwo}%

\theoremstyle{thmstylethree}%

\newcommand{\mrm}[1]{\mathrm{#1}}
\newcommand{\nuc}[2]{$\mrm{^{#2}#1}$}
\newcommand{\pos}[1]{$\mrm{e^+}$}
\newcommand{\poss}[1]{$\mrm{e^+s}$}
\newcommand{\ele}[1]{$\mrm{e^-}$}
\newcommand{\eles}[1]{$\mrm{e^-s}$}
\newcommand{\gray}[1]{$\gamma$-ray\xspace}
\newcommand{\grays}[1]{$\gamma$-rays\xspace}

\usepackage{hyperref}% if you want/for demonstration of  hyperlinks
\hypersetup{draft}
\usepackage{tablefootnote}

\usepackage{xcolor}

\begin{document}

\title[The Positron Puzzle]{The Positron Puzzle}

\author*[1]{\fnm{Thomas} \sur{Siegert}}\email{thomas.siegert@uni-wuerzburg.de}

\affil*[1]{Julius-Maximilians-Universität Würzburg, Fakultät für Physik und Astronomie, Institut für Theoretische Physik und Astrophysik, Lehrstuhl für Astronomie, Emil-Fischer-Str. 31, D-97074 Würzburg, Germany}


\abstract{The Positron Puzzle is a half-century old conundrum about the origin of the Galactic $\gamma$-ray emission line at photon energies of 511\,keV, and the shape of its morphology, showing a bulge-to-disk luminosity ratio of $\sim 1$ -- unlike any astrophysical source distribution.
Positrons (\poss{}) that have been cooled to the eV scale capture electrons (\eles{}) and form the intermediate bound state of Positronium (Ps) which decays on a nano-second timescale into two or three photons.
Assuming the emission to originate from the Galactic bulge, centre, and disk, a \emph{visible} annihilation rate in the Milky Way of $\sim 5 \times 10^{43}\,\mathrm{e^+\,s^{-1}}$ has to be explained, either by a quasi-steady state of production and annihilation, or by possibly multiple burst-like events that flood the Galaxy with \poss{}, then fading away on a Myr timescale.\\
In this paper, I will review what the real Positron Puzzle is, where data and simulations have been used inadequately which resulted in false claims and an apparent quandary, what we really know and absolutely not know about the topic, and how this epistemic problem might be advancing.\\
}




\maketitle

\section{Introduction}\label{sec:intro}
\subsection{What do we see?}\label{sec:imaging}

\begin{figure*}[t]%
	\centering
	\includegraphics[width=0.9\textwidth]{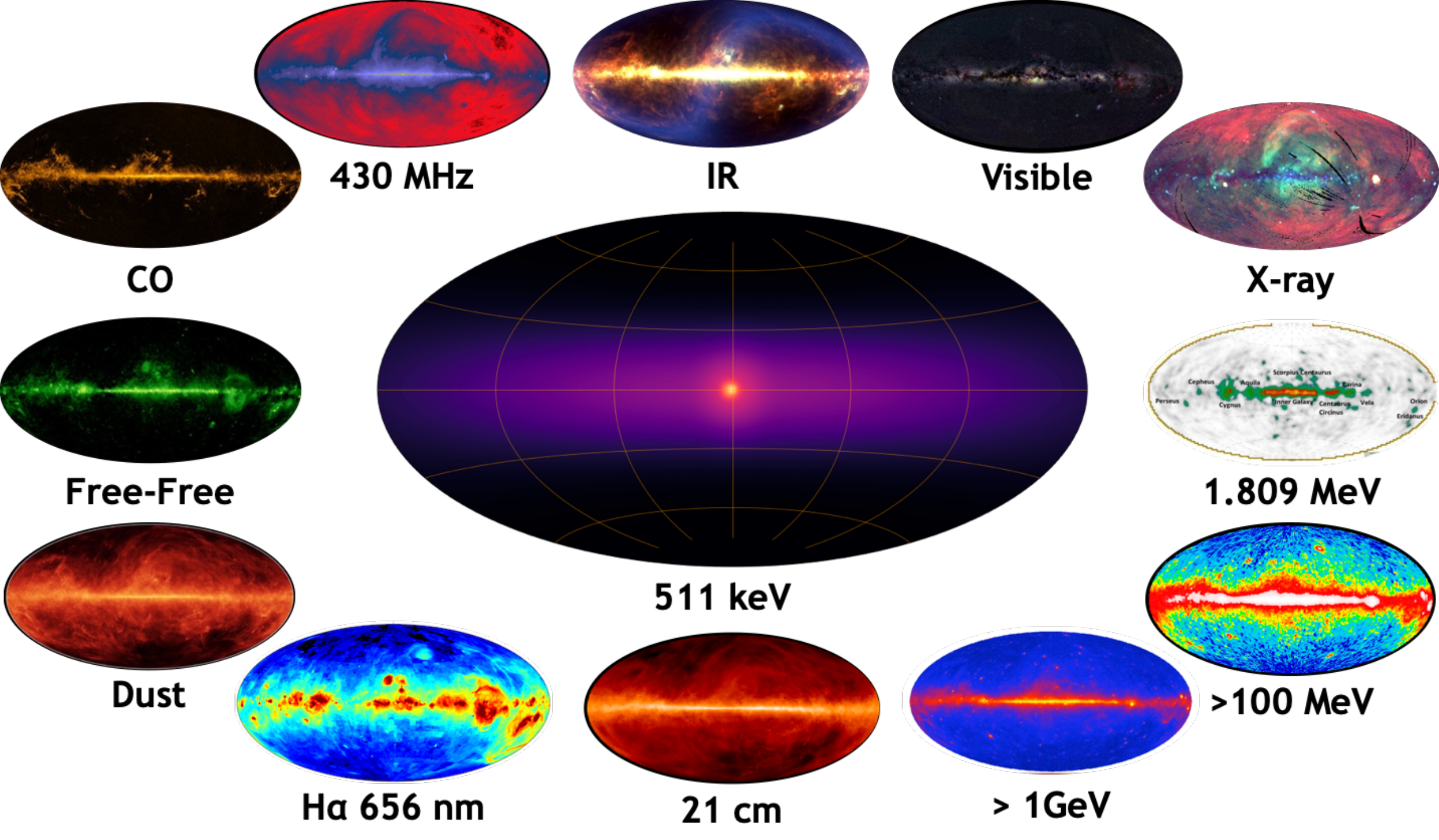}
	\caption{Full-sky maps at different wavelengths and emission processes compared to a maximum likelihood solution from INTEGRAL/SPI data at 511\,keV \citep{Siegert2016_511}.}
	\label{fig:line_map_wavelengths}
\end{figure*}

The Galactic 511\,keV signal has been measured with multiple instruments for the last 50 years.
Acknowledging the pioneering balloon experiments in the 1960s and '70s \citep{Haymes1969_511}, as well as the first identification of its `diffuse' nature with satellite experiments in the '90s \citep[CGRO/OSSE,][]{Purcell1997_511} (see also \citep{Albernhe1981_511,Lingenfelter1989_511}), the most extensive information in this topic has been gathered from INTEGRAL/SPI measurements \citep{Winkler2003_INTEGRAL,Vedrenne2003_SPI}.
With increasing exposure time since 2002, SPI measurements allowed us to refine earlier conjectures about its possible dark matter \citep[DM,][]{Boehm2004_dm} origin, conditions as to how \poss{} annihilate \citep[e.g.,][]{Jean2006_511,Churazov2005_511,Siegert2019_lv511}, and tried to infer whether, and if how far, \poss{} propagate \citep[e.g.,][]{Jean2009_511ISM,Martin2012_511,Alexis2014_511ISM,Siegert2022_511}.
The image that we can obtain with SPI at 511\,keV photon energies is unique compared to other wavelengths and emission processes (Fig.\,\ref{fig:line_map_wavelengths}).
The latest measurement of 511\,keV fluxes in the Galaxy are from \citet{Skinner2014_511} and \citet{Siegert2016_511}, finding $F_{\rm B} = (8.9$--$10.1) \times 10^{-4}\,\mrm{ph\,cm^{-2}\,s^{-1}}$, $F_{\rm D} = (13.1$--$20.1) \times 10^{-4}\,\mrm{ph\,cm^{-2}\,s^{-1}}$, and $F_{\rm GCS} = (0.6$--$1.2) \times 10^{-4}\,\mrm{ph\,cm^{-2}\,s^{-1}}$ for the Galactic bulge, disk, and centre, respectively.
The flux ratio between the bulge and disk emission is $F_{\rm B} / F_{\rm D} = 0.6 \pm 0.1$ \citep{Skinner2014_511,Siegert2016_511}, which results in a luminosity ratio of $L_{\rm B} / L_{\rm D} = 1.0 \pm 0.1$ when assuming an effective distance to the bulge of 8.2\,kpc and to the disk of 6.5\,kpc \citep{Siegert2016_511}.
Historically, the bulge-to-disk ratio was even larger because the disk was not detected in earlier measurements with SPI with less exposure time \citep[e.g.,][]{Knoedlseder2005_511,Bouchet2010_511}.
It is not excluded that the disk may actually not be \emph{the} Galactic disk, but rather halo emission or even foreground.
In the latter two cases, inferences with SPI would suffer from its coded aperture mask design, being almost incapable of measuring isotropic emission or shallow gradients \citep{Caroli1987_codedmask,Siegert2022_gammaraytelescopes}.
In addition, \citet{Siegert2019_SPIBG} showed that there is hardly a trend of Doppler-shifts in the 511\,keV line along Galactic longitudes, which would verify a rotation curve and the extended emission beyond the bulge as \emph{the} Galactic disk.
Certainly, \poss{} at the MeV scale are produced inside the Galactic disk (Sec.\,\ref{sec:astro_sources}), but how far they propagate, where they finally annihilate, and at what rate is a matter of debate (see also the works by \citet{Prantzos2006_511} and \citet{Higdon2009_511} for more details about a possible scenario to channel \poss{} from the disk to the bulge).

The central map in Fig.\,\ref{fig:line_map_wavelengths} shows a maximum likelihood fit to the raw SPI count data, requiring four components: a disk, and three components to describe the bulge, including a point-like source coincident with the Galactic centre.
Note that the angular resolution of SPI is $2.7^{\circ}$, so that this point source encompasses about 400\,pc in diameter -- reminiscent of the entire Central Molecular Zone.
The bulge component also shows an asymmetry, peaking around $l = -1^{\circ}$, which is consistently found in different analyses with different states of accumulated exposure \citep{Weidenspointner2006_511,Bouchet2010_511,Skinner2014_511,Siegert2016_511,Siegert2022_511}.
Structured, i.e. more granular images from reconstruction algorithms show a similar trend of where the flux is enhanced, and could reveal details if were it not for the strong instrumental background in MeV telescopes.
While the basic structure is also found with Richardson-Lucy \citep{Knoedlseder2005_511}, Maximum Likelihood \citep{Bouchet2010_511}, or Maximum Entropy \citep{Siegert2017_PhD} deconvolutions, image artefacts naturally emerge from the finite number of photons detected and to be distributed over a large number of pixels.
High-resolution spectroscopy of the 511\,keV line for the bulge and disk components suggests that the annihilation of \poss{} occurs dominantly in the interstellar medium (ISM), which would partly explain the `diffuse' nature of the image:

\subsection{Positron annihilation spectroscopy}\label{sec:spectra}
\begin{figure}[t]%
	\centering
	\includegraphics[width=1.0\columnwidth,trim=2.5cm 1.5cm 1.5cm 2.0cm,clip=True]{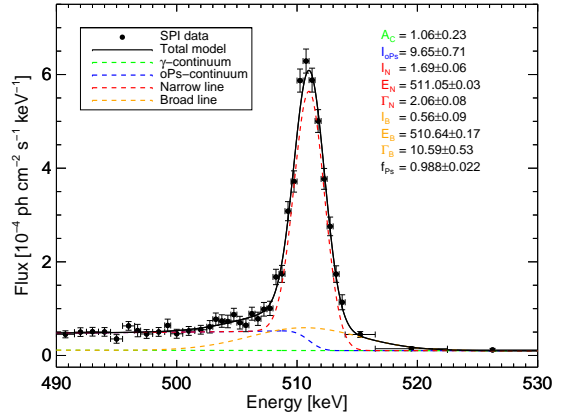}
	\caption{INTEGRAL/SPI spectrum of the entire Galaxy, decomposed into a narrow and a broad line, in addition to the ortho-Ps continuum and the Galactic diffuse continuum \citep{Siegert2019_lv511}.}
	\label{fig:line_spectrum}
\end{figure}

Previous works \citep[e.g.,][]{Churazov2005_511,Churazov2011_511,Jean2006_511,Guessoum2005_511,Guessoum2010_511,Siegert2016_511} consistently find that, assuming \poss{} annihilate in the ISM, the temperature and ionisation state of the gas in which they are annihilating is 7000--40000\,K and 2--25\,\%, respectively.
The dominant annihilation process is then charge exchange with neutral and moderately warm gas.
Charge exchange, for example with hydrogen, is only possible if the \poss{} have a kinetic energy of at least 6.8\,eV (binding energy of H minus the binding energy of Ps), which would correspond to an ISM temperature of 80000\,K.
If \poss{} reach the temperature of the ISM, they `thermalise', i.e. they are relaxing their kinetic energies to that of a Maxwellian distribution, so that the energy thresholds can be overcome only in the tail of the distribution.
One model that fits the narrow and broad 511\,keV line in the Milky Way is described as 49\,\% annihilation in the warm neutral phase and 51\,\% in the warm ionised phase \citep{Jean2006_511}:
In the ionised phase, \poss{} annihilate after thermalisation by the formation of Ps via radiative recombination with free \eles{} (no energy threshold), and to a lesser extent direct annihilation with \eles{}, forming a $\sim 1$\,keV broad line (FWHM), without any Ps formation in flight.
In the warm phase, most \poss{} form Ps in flight, resulting in a $\gtrsim 6$\,keV broad line, and a small percentage thermalises, again forming Ps, now with a narrow $\sim 1$\,keV line.
The total Ps fraction,
\begin{equation}
	f_{\rm Ps} =  \frac{8r_{32}}{9 + 6r_{32}}\mrm{,}
	\label{eq:Ps_fraction}
\end{equation}
with $r_{32} = F_{\rm oPs}/F_{511}$ being the flux ratio between the ortho-Ps continuum and the 511\,keV line, describes the fraction of \poss{} that undergo the formation of Ps \citep[e.g.,][]{Leventhal1978_511}.
In the warm ionised medium $f_{\rm Ps}$ is $\sim $88\,\% and in the warm neutral medium $\sim 100$\,\%, resulting in a total Ps fraction across the Galaxy of $92$--$97$\,\%.
In fact, some of the broadening of the Galactic 511\,keV line may also be due to velocity broadening from Galactic rotation \citep{Siegert2019_lv511}.
Furthermore, the line is not the only witness of annihilating \poss{}: the ortho-Ps component \citep{Ore1949_511} is also clearly visible in the combined spectra across the Galaxy, rising almost linearly with a sharp cut off at 511\,keV, and containing a $\approx 4$ times larger annihilation flux, depending on $f_{\rm Ps}$ (Fig.\,\ref{fig:line_spectrum}).

All these assumptions make perfect sense, if the premise of `annihilation in the ISM' is correct.
However, the density of target particles also has a considerable impact on the line shape and Ps fraction, once the densities of the gas is much higher than that of the ISM:
\citet{Murphy2005_posiloss} calculate the Ps fraction as a function of temperature and gas density in the context of solar flares, finding that $\sim 1$\,keV broad 511\,keV lines can be explained in gas with densities $\gtrsim 10^9\,\mrm{cm^{-3}}$ and temperatures $\lesssim 2 \times 10^4$\,K.
Also the broad- to narrow-line ratio is found to be below $2$ for temperatures above $\sim 8000$\,K.
Finally, the Ps fraction, albeit on the lower side, can also be met in conditions typical to solar flares.

\subsection{A puzzling picture}\label{sec:puzzle}
The `Positron Puzzle' may be summarised in three basic questions:
\begin{enumerate}
	\item What do we see?
	\item Where do the positrons come from?
	\item Why does it look like that?
\end{enumerate}
The answer to the first question, albeit somewhat ambiguous considering the limitations of current instrumentation in the MeV $\gamma$-ray band, is outlined in Secs.\,\ref{sec:imaging} and \ref{sec:spectra}.
Measurements in the MeV regime suffer from a scarcity of photons from astrophysical sources, carrying a million times more energy in one quantum than optical photons, for example, large instrumental background \citep{Diehl2018_BGRDB,Siegert2019_SPIBG}, and apertures which align more with particle detectors in space rather than telescopes \citep{Siegert2022_gammaraytelescopes}.
Images are therefore often either `reconstructed', i.e. it is attempted to unfold the non-invertible imaging response matrices of MeV instruments, or models are directly fitted to the raw data.
In performing these imaging techniques, however, comparisons to other wavelengths, that could be more directly measured, can result in false conclusions.
But in addition to the imaging capabilities, also the spectroscopic information of the 511\,keV line will help to decipher `how' the \poss{} are annihilating.
In the paragraphs in Sec.\,\ref{sec:tracers}, I outline possible tracers of \pos{}-annihilation, based on first, the visual comparison of all-sky surveys from Fig.\,\ref{fig:line_map_wavelengths}, and second, a thorough likelihood comparison.

The second question is the historically initial question after \citet{Leventhal1978_511} unambiguously identified the $\gamma$-ray line from the Galactic centre as due to \pos{}-annihilation.
``\textit{Undoubtedly, the positrons giving rise to the observed feature come from a variety of processes}'', said \citet{Leventhal1978_511} about this first and strongest $\gamma$-ray line detected.
Indeed, where the \poss{} come from, i.e. in which astrophysical sources they originate, is related to how \poss{} can be produced in terms of particle physics processes.
I will summarise possible \pos{} sources and production mechanisms in Sec.\,\ref{sec:sources}, together with measurements of how much which source population might contribute to the Galactic \pos{} budget.

But the budget alone is not enough to gain complete insight into the `Positron Puzzle' because the sources of \poss{} may in fact not be the sinks of \pos{} annihilation.
This is related to the third question of why the emission looks so different than the putative sources, and also different to the supposedly preferred annihilation regions of dense gas in the Milky Way.
Positrons are typically produced at mildly- or highly-relativistic energies (see Sec.\,\ref{sec:sources}).
This means, unless they experience favouring annihilation conditions, such as high gas or \ele{} densities and low temperatures, they \emph{are} cosmic rays, and propagate through the ISM.
In Sec.\,\ref{sec:propagation}, I will summarise the current knowledge of propagation of \poss{} in the ISM, which links to the distortion of the potential source distributions towards the measured 511\,keV image.
The injection of \eles{} and \poss{} into the ISM will transport the particles on timescales of 0.1–10\,Myr to distances (path lengths) of 0.1–10\,kpc away from their initial sources \citep[e.g.,][]{Jean2009_511ISM,Panther2018_pos_transport}.
Consequently, to first order, the 511\,keV image could be a smeared-out version of the initial source distribution due to the propagation effects, which may be similar to the gas distribution of the Galaxy.

The Milky Way can either be described in terms of its stellar population (putative \pos{} sources) or in terms of its gas content (possible \pos{} annihilation regions).
The stellar population can be subdivided in the Galactic bulge and the Galactic disk \citep{Freudenreich1998_BoxyBulge_COBE}.
The bulge shows a scale size of $\sim 1.6$\,kpc, an old stellar population ($\sim 10$\,Gyr), and a stellar mass of $\sim 1.5 \times 10^{10}\,\mrm{M_{\odot}}$ .
The disk can be divided further into a thin (young(er) stars, on average $\sim 5$\,Gyr) with scale height and radius of $\sim 300$\,pc and $2.6$\,kpc, respectively, and a thick disk (old stars, $\sim 10$\,Gyr) with scale heigh and radius of $900$\,pc and $3.6$\,kpc, respectively.
The stellar masses of the thin and thick disk are $\sim 2.5$ and $\sim 0.5 \times 10^{10}\,\mrm{M_{\odot}}$, respectively.
The gas in the Galaxy can be roughly subdivided into a molecular, an atomic, and an ionised component, with scale heights of $\sim 50$\,pc \citep{Dame2001_COsurvey}, $\sim 100$\,pc \citep{Dickey1990_HI}, and $\sim 1800$\,pc \citep{Gaensler2008_ionised_gas}, respectively.
The total gas mass of the disk (2--20\,kpc) is on the order of $10^{10}\,\mrm{M_{\odot}}$, similar to that of the bulge gas mass \citep{Ferriere2001_ISM,Ferriere2007_ISMbulge}.
In addition to the bulge and disk gas reservoirs, there is also the Central Molecular Zone \citep[CMZ,][]{Morris1996_GC} encompassing the Galactic centre.
It is $\sim 400$\,pc in size with a strong concentration of molecular gas.

\subsection{Positron annihilation tracers}\label{sec:tracers}
With the information above, we can investigate Fig.\,\ref{fig:line_map_wavelengths} for tracers of \pos{} annihilation:
\paragraph{Ionisation:} Ionisation and recombination of hydrogen \citep{Seaton1959_radrecomb}, as would happen as a consequence of the charge exchange process and radiative recombination with \eles{}, may be traced in the ISM by H$\alpha$ in the Milky Way at photon wavelengths of 656\,nm \citep{Finkbeiner2003_Halpha}.
The H$\alpha$ emission appears clumpy with several hotspots along the Galactic plane as well as nearby star forming regions, such as Orion, Scorpius Centaurus, Cygnus, Carina, or Vela.
There is no definite bulge emission that would outshine the disk, and the bulge-to-disk ratio is tiny.
\paragraph{Warm ISM:} The distribution of neutral and partly warm gas in the Milky Way is visible in the 21\,cm line of atomic hydrogen \citep[HI; e.g.,][]{Kerp2011_HI}.
The emission is concentrated in a thin disk along the Galactic plane with visible cavities, so-called superbubbles \citep{Weaver77,Castor75}, that have been formed by massive star winds and supernovae (SNe) over Myr timescales.
The cavities may extend to several kpc above the Galactic plane so that the total HI emission may be described by a thin and a thick disk.
The map shows where the gas might be found for the \poss{} to annihilate in, but there is no bright bulge at 21\,cm wavelengths.
\paragraph{Cold ISM:} Based on logic arguments and reasonable estimates for the lifetime of \poss{} in the ISM \citep{Panther2018_alkali511}, the higher the density of the ISM, the more \poss{} should be seen to annihilate.
The CO map \citep[$J = 1 \rightarrow 0$ transition,][]{Dame2001_COsurvey} at 115\,GHz shows the dense and cold gas -- regions where \poss{} should naturally tend to reach for annihilation if not quenched by too low temperatures to undergo charge exchange.
The thin disk of CO is nothing like the 511\,keV map, and also no bulge is present.
However, the CMZ would be coincident with the point-like source at 511\,keV.
\paragraph{Dust grains:} It appears that annihilation on dust grains is not excluded by the spectral data although the emission morphology, again, would be disfavoured by direct comparison of the dust map \citep[e.g.,][]{Bennett2013_WMAP,Planck2016_foregrounds} in Fig.\,\ref{fig:line_map_wavelengths} to the 511\,keV map.
Also PAHs (polycyclilc aromatic hydrocarbon molecules) may be traced by dust emission, and would make excellent targets for \poss{} to annihilate with because of their huge `effective charge number', $Z_{\mathrm{eff}}$, and therefore reaction cross section.
PAHs show $Z_{\mathrm{eff}}$ values of up to $10^8$, so that their cumulative effect in the ISM may lead to reaction rates of $\sim 10$\,\% of the dominating processes in the warm neutral and ionised medium, even though their number abundance is only $10^{-6}$ \citep{Guessoum2010_511}.
It is apparent that the Milky Way dust map does not trace the 511\,keV map, except again for the possible high latitude emission.
\paragraph{Particle acceleration and propagation:} Annihilation of \poss{} in flight \citep[see Sec.\,\ref{sec:propagation}, and, e.g.,][]{Beacom2006_511,Sizun2006_511}, either through charge exchange or directly, may be traced by several processes that should show their paths of transport.
Positrons should be bound to the large-scale magnetic field of the Galaxy, so that the synchrotron emission \citep[e.g. at 408\,MHz,][]{Haslam1981_408MHza} could show places where \poss{} annihilate.
Likewise, bremsstrahlung \citep[free-free emission,][]{Planck2016_Dust} could possibly show the catastrophic energy loss of \poss{}, being removed from the scheme due to annihilation in flight.
Electron bremsstrahlung and \pos{} bremsstrahlung are hardly distinguishable, so that the free-free emission map in Fig.\,\ref{fig:line_map_wavelengths} may show both particle populations.
At very high energies, beyond 100\,MeV, the Galactic emission is dominated by pion decay \citep[$\pi^0 \longrightarrow 2\gamma$,][]{Bjorklund1950_piondecay} as a result of high-energy cosmic-ray interactions with the dense ISM \citep[e.g.,][]{Ackermann2013_CR}.
Therefore, this emission might represent both, production site and destruction site of \poss{}.
But also for pion decay, the bulge emission would be missing as seen in the Fermi/LAT image above $1$\,GeV in Fig.\,\ref{fig:line_map_wavelengths}.
At these photon energies, there is also no high-latitude component\footnote{The Fermi Bubbles \citep[e.g.,][]{Su2010_fermibubbles} are high-latitude emission but confined to longitudes $-10^{\circ} \lesssim \ell \lesssim 10^{\circ}$.} and the emission is clustered around star-forming regions, dense molecular clouds, and spiral arms.
\paragraph{In-situ annihilation:} If the propagation of \poss{} is not ballistic (Sec.\,\ref{sec:propagation}), i.e. not determined by collisional transport \citep{Jean2009_511ISM}, they might be advected with the Galactic velocity field rather than diffusing several kpc.
This may result in morphologies that would resemble the possible sources of \poss{} ( Sec.\,\ref{sec:sources}), such as nucleosynthesis ejecta (e.g. shown as the 1.809\,MeV map), X-ray binaries (XRBs; e.g. shown as the X-ray all-sky map from ROSAT), or, indeed, stars themselves through flares \citep[traced by infrared and visible light,][]{Bisnovatyi-Kogan2017_511}.
Especially in the latter case, the need for \pos{} propagation in the ISM would be weakened as parts of the emission are entirely explained so that a different set of constraints will emerge.
This may include only propagation inside the bulge, for example, and no channelling of \poss{} from the disk into the bulge \citep[e.g.,][]{Higdon2009_511}.
For XRBs, a population of point-like sources would be expected for the 511\,keV emission, or, after propagation, again an image similar to the HI, CO, or dust maps.
The hot tenuous X-ray gas with temperatures above 1\,MK is probably too thin for \poss{} to efficiently cool down and annihilate in large numbers (see, however, Sec.\,\ref{sec:low-energy-propagation}).
The 1.809\,MeV emission from \nuc{Al}{26}, a $\beta^+$-decaying massive star nucleosynthesis product, shows potentially the negative image of HI, i.e. instead of the shell boundaries now the interiors.
These cavities are filled with hot gas streaming away from the stellar groups into lower pressure regions \citep{Krause2015_26Al}.
Again, \poss{} would start out at low densities and high temperatures, so that only after reaching the HI walls, they would annihilate, then in the vicinity of the sources.
More information about the possible sources of Galactic \poss{} are presented in Sec.\,\ref{sec:sources}.

\begin{figure}[t]%
	\centering
	\includegraphics[width=1.0\columnwidth]{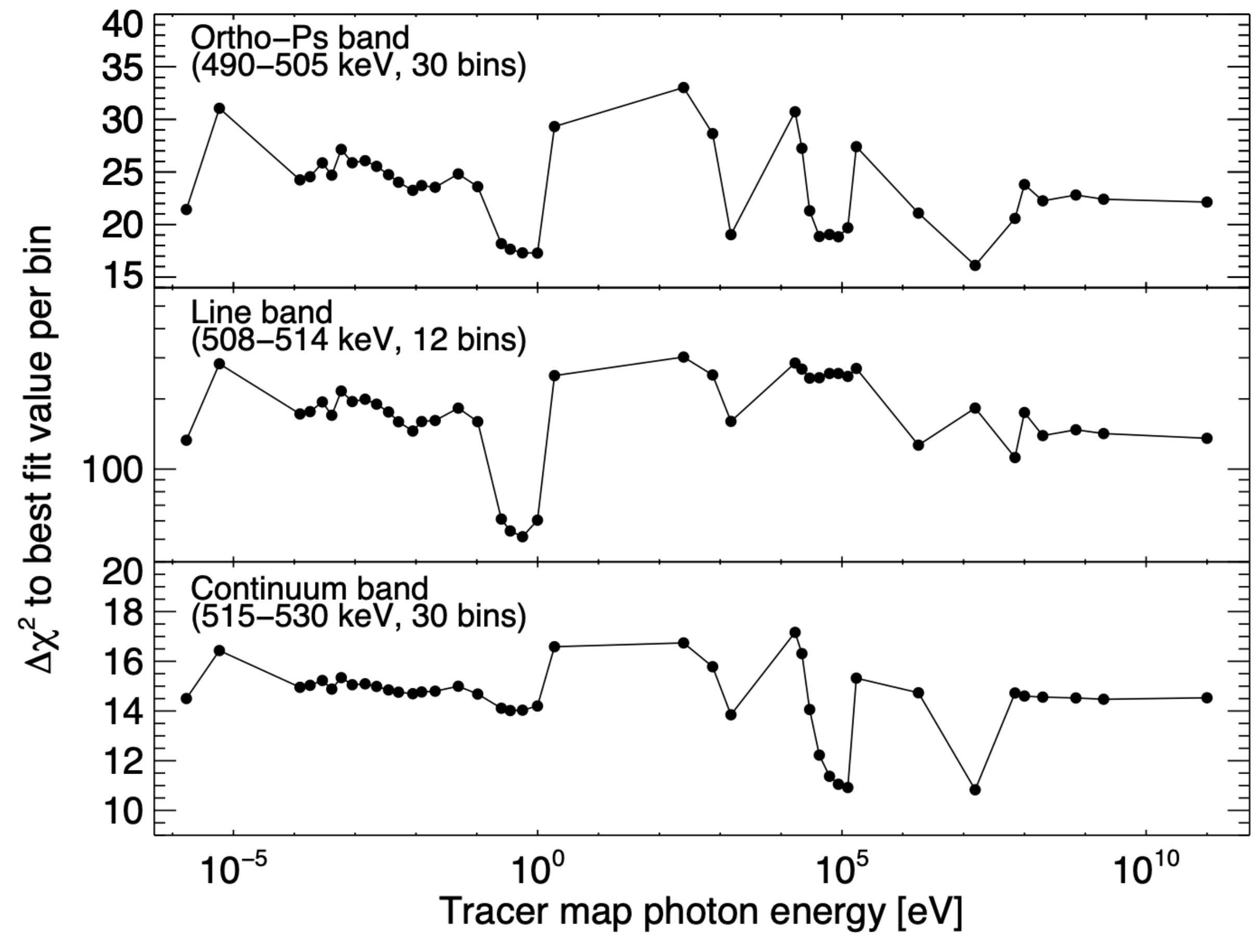}
	\caption{Likelihood comparisons of different tracer maps as a function of photon energy in three bands, including the ortho-Ps continuum, the 511\,keV line, and the diffuse Galactic continuum emission \citep{Siegert2017_PhD}.}
	\label{fig:likelihood_comparisons}
\end{figure}

\paragraph{Statistical comparison:} 
The visual inspection of these maps with the physical arguments from spectroscopy should not be over-interpreted because the reconstruction of the 511\,keV emission itself is biased by the assumptions of how a Galaxy may look like.
Instead, one can define an adequacy-of-fit criterion\footnote{This is not a goodness-of-fit criterion, since it is not statistically sound to judge a fit based on such a likelihood comparison.}, $\Delta\chi^2 = -2 \left[ \ln\mathscr{L}(\mathrm{BG+SKY}) - \ln\mathscr{L}(\mathrm{BG})\right]$.
This may be interpreted as a relative measure, which template maps, either by wavelength or physical process, fit the raw data best.
I want to stress, that this is not an absolute measure of how good the fit is, nor is this intended.
Fig.\,\ref{fig:likelihood_comparisons}, shows the $\Delta\chi^2$-values as a function of tracer map photon energy for the ortho-Ps band, the 511\,keV line band, and the continuum band above the \pos{}-annihilation line \citep{Siegert2017_PhD}.
It can be seen that the line and ortho-Ps behave similarly as a function of tracer map photon energy, except for the 10--200\,keV range.
This is understandable because the continuum emission above the line is nicely traced by X-ray sources as seen by the dip (improvement) in $\Delta\chi^2$ in the lower panel, and also impacts the emission below the line.
The only all-sky maps that consistently show an improvement in $\Delta\chi^2$ for both 511\,keV and ortho-Ps are the infrared maps below 1\,eV ($1.25$--$4.9\,\mathrm{\mu m}$) from COBE/DIRBE.
This improvement is not seen in the continuum band, so that it is prone to \pos{}-annihilation.

This means that star light (and the zodiacal light, which has not been removed from the maps; see also Sec.\,\ref{sec:foreground}) apparently traces annihilating \poss{} -- not their presumed sources, nor the ISM.
Stars may be both, the dominant sources and the natural sinks, of \poss{} in the Galaxy.
In fact, contemporaneous to \citet{Siegert2017_PhD}, \citet{Bisnovatyi-Kogan2017_511} suggested flaring stars to be a major contributor to the Galactic 511\,keV signal.

Such a template-fit analysis has also been performed by \citet{Knoedlseder2005_511} to the 511\,keV line, with only 1.5\,yr of exposure in the Galaxy, resulting in similar conclusions that the emission might be related to the or an old stellar population \citep[see also][]{Siegert2022_511}.
These types of analyses may appear odd for non-$\gamma$-ray astronomers, but because of the fact that we cannot directly (or only hardly) image the $\gamma$-ray sky leads us to such fundamental comparisons.
\citet{Knoedlseder1999_freefree1.8MeV} performed also a similar study for the origin of the 1.8\,MeV line from decaying \nuc{Al}{26}, finding that the free-free emission at 53\,GHz is correlating with the measurements from COMPTEL.
This result was part of the solid proof that massive stars are indeed responsible for the observed \nuc{Al}{26}.
The conjecture for the \pos{} sources is `\emph{the old stellar population}', but what exactly this entails is not yet clear, since the interpretations are still versatile.

\section{Positron sources}\label{sec:sources}
The general, straight-forward, question of the Positron Puzzle is `\emph{Where do the positrons come from?}', meaning, what are the sources to explain the amount that annihilates in the Milky Way.
The list of potential candidate sources is large because nearly every astrophysical source can produce \poss{} either directly or through secondary interactions with the surroundings.
Consequently, the actual problem is not \emph{which} sources are responsible, but rather that there are too many possibilities\footnote{Quoted after a lecture by G.~K.~Skinner.}.

Another fundamental predicament of the Positron Puzzle is that it appears that its solution will only benefit the puzzle itself, then being solved.
But the fact that nearly every astrophysical source indeed is able to produce (and annihilate) \poss{} may turn the 511\,keV emission line into an independent astrophysical messenger, a tool to study other unsolved problems, a way to gain a perspective orthogonal to other measurable quantities.
When the puzzle is solved, and it eventually turns out that many different astrophysical source types contribute to the Galactic 511\,keV emission, a dedicated observatory may revolutionise our understanding in the multiple fields of astrophysics associated with it.
In the following, I will summarise candidate sources and put them into the context of available observations to estimate their contribution, given measurement uncertainties and systematics from model assumptions.
This is closely following \citet{Siegert2017_PhD} but will be more concise.
For completeness, I will briefly describe the particle physics processes that can lead to \poss{}.

\subsection{Positron production mechanisms}\label{sec:production}
\paragraph{$\beta^+$-decay:} The most intuitive answer to the question where \poss{} could emerge from is the radioactive decay of isotopes with a proton excess.
The generic equation for $\beta^+$-decay is
\begin{equation}
	\mathrm{^A_{Z}X \stackrel{\tau_{A,X}}{\longrightarrow} {}^A_{Z-1}Y + e^+ + \nu_e,}
	\label{eq:betaplusdecay}
\end{equation}
where a nucleus $\mathrm{X}$ with charge number $Z$ and atomic number $A$ decays within its lifetime of $\tau_{A,X}$ to another nucleus $\mathrm{Y}$ with reduced charge number $Z-1$ but same atomic number by the emission of a \pos{} and an electron-neutrino $\mrm{\nu_e}$.
For most nuclei, this is not the sole decay branch so that it is associated with a branching ratio, or equivalently, a probability $p_{\beta +}$ for this path.
Astrophysically important nuclei can be associated to different objects (see Sec.\,\ref{sec:astro_sources}) and are listed in Tab.\,\ref{tab:positron_emitters} for completeness.
It becomes clear that there is a variety of eleven orders of magnitude in lifetime of these relevant isotopes.
This means \poss{} may either annihilate far away from their initial production sites, or they cool down quickly due to a possibly large density and find \eles{} to recombine with at the position of their sources.
A mixture of \pos{} annihilation after propagation and in-situ would be the result.
\begin{table*}[t]
	\begin{center}
			\begin{minipage}{\textwidth}
				\centering
			\caption{List of astrophysically important positron emitting nuclei, sorted by lifetime $\tau$. The columns are the nucleus, its lifetime, the probability to emit a positron while decaying, possibly associated $\gamma$-ray emission from the daughter nucleus in units of MeV, and potential sources.}\label{tab:positron_emitters}%
			\begin{tabular}{l c r p{3cm} p{6cm}}
				\toprule
				Nucleus & $\tau$ & $p_{\beta+}$ & $E_{\gamma}$ & Sources \\
				\midrule
				\nuc{Al}{26} & 1.03\,Myr & 0.82 & 1.809 & Massive stars, AGB stars, Supernovae \\
				\nuc{Sc}{44} & 81\,yr$^a$ & 0.94 & 1.157 & Supernovae \\
				\nuc{V}{48}$^d$ & 23.1\,yr & 0.50 & 0.983, 1.312 & Supernovae \\
				\nuc{Na}{22} & 3.75\,yr & 0.90 & 1.275 & Novae \\
				\nuc{Co}{56} & 111.4\,d$^b$ & 0.20 & 0.847, 1.238 & Supernovae \\
				\nuc{Ni}{57}$^d$ & 2.14\,d & 0.43 & 0.127, 1.378, 1.920, 0.122$^c$, 0.136$^c$ & Supernovae \\
				\nuc{F}{18} & 2.64\,h & 0.97 & -- & Novae, Solar flares \\
				\nuc{Mn}{52}$^d$ & 30.4\,min & 0.29 & 0.744, 0.936 & Supernovae \\
				\nuc{C}{11}$^d$ & 29.3\,min & $>$0.99 & -- & Cosmogenic (cosmic-ray interactions, spallation), Solar flares \\
				\nuc{N}{13} & 14.4\,min & $>$0.99 & -- & Novae, Earth atmosphere / lightning, Solar flares \\
				\nuc{O}{15} & 2.94\,min & $>$0.99 & -- & Novae, Earth atmosphere / lightning, Solar flares \\
				\botrule
			\end{tabular}
	%\vspace{-1.5ex}
	\end{minipage}
	\begin{minipage}{\textwidth}
	\footnotesize{$^a$ The nucleus \nuc{Sc}{44} only has a half-life time of 3.9\,h and exists only as an intermediate step from the decay of \nuc{Ti}{44}. The relevant astrophysical timescale, for example for heating of supernova remnants, is that of the longer-living \nuc{Ti}{44}.
		
		$^b$ The nucleus \nuc{Co}{56} is the daughter product of the shorter-lived \nuc{Ni}{56} that is dominantly produced in supernovae. The relevant timescale here is again that of the longer-living \nuc{Co}{56}.
		
		$^c$ The $\gamma$-rays at 122 and 136\,keV come from the daughter nucleus' decay, \nuc{Co}{57}$\rightarrow$\nuc{Fe}{57} which is no $\beta^+$-decay, but the $\gamma$-rays might indicate that positrons have been emitted throughout the \nuc{Ni}{57} decay chain.
		
		$^d$ These isotopes have not been considered for the Positron Puzzle so far but may play a role.}
	\end{minipage}
	\end{center}
\end{table*}

\paragraph{Particle acceleration:} Positively charged mesons, such as pions and kaons, may also emit $\mrm{e^+}s$, formally involving the same process of $\beta^+$-decay.
However, their production is only possible from the interaction of high-energy particles (including photons).
Hadronic processes of the shape
\begin{equation}
	\mrm{p + p \longrightarrow \pi^{0,+,-} + X,}
	\label{eq:pion_production}
\end{equation}
and
\begin{equation}
	\mrm{p + \gamma \longrightarrow n + \pi^+, p + \pi^0}
	\label{eq:photopion_production}
\end{equation}
are possible for threshold energies above $\sim 280$\,MeV ($m_{\pi0} \approx 135$\,MeV, $m_{\pi\pm} \approx 140$\,MeV).
For pp-interactions, Eq.\,(\ref{eq:pion_production}), the production cross section rises logarithmically, reaches a plateau on the order of $1.4 \times 10^{-25}\,\mrm{cm^2}$ above $\sim 1$\,TeV, and may rise afterwards again.
Each $\pi^+$ produced will eventually lead to one \pos{}, either directly or through the decay of an anti-muon.
Because the pions produced will have large kinetic energies, the $\mrm{e^+}s$ will carry away this energy with Lorentz factors of $10^3$ and more, but with a peak around 30--40\,MeV \citep{Murphy1987_pionpositrons}.
Cosmic-ray production of secondary \poss{} will therefore lead to a very smeared out annihilation morphology as the \poss{} may be prone to large-scale propagation.
For photo-pion production, Eq.\,(\ref{eq:photopion_production}), the arguments are similar, but the interaction cross section is reduced to $\sim 10^{-28}\,\mrm{cm^2}$.
Neutral pions have a dominant decay branch into two photons with 98.8\,\%; the remaining $\sim 1.2$\,\% allow for a decay into one photon and an \ele{}-\pos{}-pair.
The effective production rate of \poss{} from neutral pions is then about 1\,\% of that of positively charged pions.

In addition to meson production, cosmic rays also lead to spallation reactions on ISM nuclei, thereby forming $\beta^+$-unstable isotopes, such as \nuc{C}{11}.

\paragraph{Pair production:} Electron-positron pairs can be created by the interaction of high-energy photons in the electric field of a nucleus (or \ele{}), in a magnetic field, or by photon-photon pair production.
The first reaction has a cross section proportional to $Z^2$, and rises logarithmically with photon energy, showing typical values of $>10^{-27}\,\mrm{cm^{-2}}$ above its threshold energy of 1.022\,MeV \citep{Maximon1962_PP}.
The second requires strong magnetic fields on the order of $10^{12}$\,G to show an efficient photon to pair conversion.
One cannot ascribe this process a cross section but rather an attenuation coefficient $\gtrsim 10^2\,\mrm{cm^{-1}}$ for a magnetic field of $5 \times 10^{12}$\,G for photon energies above the threshold.
Finally, photon-photon pair production has a maximum cross section close to the Thomson cross section of $0.25\sigma_T = 1.66 \times 10^{-25}\,\mrm{cm^{2}}$ at a centre of momentum energy of $4 \times 511$\,keV.
The cross section is suppressed for energies near the threshold and decreases $\propto \gamma^{-1}$ for higher energies.
Photon-photon absorption is important in the context of the mean free path, or, equivalently, the optical depth of very high-energy photons, for example from AGN or Galactic `pevatrons':
High-energy photons interact with the ubiquitous photon field of the cosmic microwave background (CMB) or the interstellar and intergalactic radiation field at infrared to optical wavelengths, leading to a cutoff in the TeV regime due to pair production.
The rate of pair production in the intergalactic medium (IGM) due to photon-photon absorption from TeV emitting AGN has, to the knowledge of the author, not yet been calculated/estimated.

\citet{Svensson1982_plasma} studied the bahaviour of a thermal proton-electron-positron-photon plasma, including the radation mechanisms of bremsstrahlung, Inverse Compton (IC) scattering, pair annihilation and pair production through all the above-mentioned processes (except magnetic fields).
This is particularly important in the case of black holes (BHs) with accretion disks, forming photon spectra up to MeV energies and therefore leading to an $\mrm{e^\pm}$ atmosphere that may be carried away in an optically thin wind \citep{Beloborodov1999_511}.
This pair-plasma would then annihilate according to its plasma temperature \citep{Svensson1982_ann_spec,Svensson1983_TPA}.
It should be noted that pair-plasma annihilation will \emph{not} produce a narrow 511\,keV line unless the plasma temperature is particularly low.
Instead, a blueshifted (Doppler-boosted), and broadened feature will emerge -- searches for narrow annihilation lines in sources where pair-plasma would be expected will therefore always be negative even though pair-annihilation may have occurred.

\paragraph{Hawking radiation:} The evaporation of BHs due to Hawking radiation is the suggested cooling (and therefore mass loss) mechanism of these compact objects \citep[e.g.,][]{Hawking1975_HawkingRadiation}.
Hawking radiation is believed to be a quasi-thermal emission mechanism with a temperature $T_{\rm BH} = 1.06 \left(10^{16}\,\mrm{g}/M_{\rm BH}\right)$\,MeV, with $M_{\rm BH}$ being the mass of the BH, of not only photons but also particles -- and therefore also \poss{}.
The kinetic energies of the particles will also have a blackbody distribution so that a large variety of energies are possible according to the range of BH masses.
Primordial black holes \citep[PBHs,][]{Hawking1971_PBH} may have masses on the order of $\gtrsim 4 \times 10^{17}$\,g \citep[e.g.,][]{Siegert2022_RetII,Berteaud2022_SPI_PBH,Iguaz2021_CGBPBH} which would result in average \pos{} energies of $\lesssim 100$\,keV, potentially not propagating far from the sources.
The associated $\gamma$-ray signature would provide a smoking-gun evidence that PBHs may be responsible for the DM phenomenon as well as a major contributor to the annihilating \poss{}.
It should be noted that Hawking radiation is not proven by observations to actually exist.

\paragraph{Dark matter:} If DM is composed of beyond-standard-model particles, their decay \citep[e.g.,][]{Hooper2004_dm}, annihilation \citep[e.g.,][]{Boehm2004_dm}, or excitation \citep[e.g.,][]{Finkbeiner2007_dm511deex} might result in \ele{}-\pos{}-pairs.
Depending on the actual particle candidate, i.e. its mass, \poss{} carry away the parent particles' rest mass energy as kinetic energy.
Thus, \poss{} may annihilate close to their production sites of light DM particles, leading to a 511\,keV emission reminiscent of a DM halo, or far away because they diffuse over large distances from GeV--TeV mass scale WIMPs.
In the latter case, the annihilation branch into massive quark-antiquark pairs is dominating over \ele{}-\pos{}-pairs \citep{Jungman1996_susydm}.
However, the intermediate formation of charged mesons will eventually lead to \poss{}, and potentially more than one per WIMP co-annihilation.
It should also be noted here that there is no observational evidence of particle DM.

\subsection{Astrophysical sources}\label{sec:astro_sources}
Given the production mechanisms, Sec.\,\ref{sec:production}, astrophysical sources can be associated, and, if observations are available, \poss{} from individual candidate populations can be budgetised.
Such an approach does not take into account the possible propagation of \poss{} (Sec.\,\ref{sec:propagation}) and shows that the puzzle is indeed either under- or over-explaining the annihilation rate in the Milky Way.
Tab.\,\ref{tab:source_rates} summarises the rates at which candidate populations (see following paragraphs) produce \poss{}, including uncertainties from measurements as well as systematics from unknown population sizes, for example.

\begin{table*}[!ht]%
	\large
	\centering
	\begin{tabular}{lccc}
		\hline
		Source & Galaxy & Bulge & Disk \\
		\hline
		$^{26}$Al & $3.4\pm0.1$ & $0.35\pm0.01$ & $3.0\pm0.1$ \\
		$^{44}$Ti & $2.2\pm0.3$ & $0.22\pm0.03$ & $1.9\pm0.3$ \\
		$^{56}$Ni & $19_{-19}^{+119}$ & $4.6_{-4.6}^{+29}$ & $14_{-14}^{+90}$ \\
		$\gamma\gamma$ & $27_{-27}^{+29}$ & $8.9_{-8.9}^{+9.5}$ & $18_{-18}^{+19}$ \\
		\hline
		$\sum$ Sources & $51_{-33}^{+122}$ & $14_{-10}^{+30}$ & $37_{-23}^{+92}$ \\
		\hline
		\vspace{5pt}
		Target Rate & $49\pm15$ & $18.0\pm0.2$ & $31\pm15$ \\
		\hline
		\hline
		& Galaxy \% & Bulge \% & Disk \% \\
		\hline
		$^{26}$Al & $7\pm2$ & $1.9\pm0.2$ & $10\pm5$ \\
		$^{44}$Ti & $4\pm2$ & $1.2\pm0.2$ & $6\pm3$ \\
		$^{56}$Ni & $39_{-39}^{+243}$ & $27_{-27}^{+160}$ & $47_{-47}^{+291}$ \\
		$\gamma\gamma$ & $55_{-55}^{+61}$ & $49_{-49}^{+53}$ & $57_{-57}^{+67}$ \\
		\hline
		$\sum$ Sources & $105_{-67}^{+251}$ & $78_{-56}^{+169}$ & $120_{-74}^{+300}$ \\
		\hline
		Target \% & $100\pm31$ & $100\pm11$ & $100\pm48$ \\
		\hline
	\end{tabular}
	\caption{Comparison between measured \poss{} annihilation and production rate (top, in units of $10^{42}\,\mrm{e^+\,s^{-1}}$) in the Milky Way. The contributions in per cent (bottom) to bulge and disk have been estimated from the ccSNe rates in the case of $^{26}$Al and $^{44}$Ti (10\,\% bulge, 90\,\% disk), from the SNe Ia rate in the case of $^{56}$Ni (25\,\% bulge, 75\,\% disk), and from the stellar distribution in the case of $\gamma\gamma$ (33.3\% bulge, 66.7\% disk). For details, see \citet{Siegert2017_PhD}.}
	\label{tab:source_rates}
\end{table*}

\paragraph{Massive stars and core-collapse supernovae:} 
Through massive star winds and their ultimate core-collapse supernovae (ccSNe), the ISM is enriched by freshly produced elements.
For the production of \poss{}, the most important isotopes in stellar evolution are probably \nuc{Al}{26}, \nuc{Ti}{44}, and \nuc{Co}{56}.
To a lesser extent also \nuc{Ni}{57}, \nuc{Mn}{52}, and \nuc{V}{48} may play a role, but have so far not been considered in the context of the Positron Puzzle, probably because of their small yields and shorter lifetimes \citep[see, however,][]{Panther2021_48V}.
The \pos{}-production rate follows directly from the radioactive decay law:
If a mass $M_i$ of radioactive, $\beta^+$-unstable, isotope $i$ is produced, the number of particles that decay per time is $N_i(t) = \frac{M_i}{m_i}p_i e^{-t/\tau_i}$, with $m_i$ being the isotope mass, $\tau_i$ its lifetime, and $p_i$ the probability of undergoing \pos{}-emission (branching ratio).
The \pos{}-production rate, $\dot{N}_{+}(t)$, from isotope $i$ is then
\begin{equation}
	\dot{N}^{+}_i(t) = - \frac{dN_i(t)}{dt} = \frac{M_i p_i}{m_i \tau_i}e^{-t/\tau_i} = \frac{N_{0,i}^+}{\tau_i}e^{-t/\tau_i}\mrm{.}
	\label{eq:positron_production_betaplus}
\end{equation}

\noindent It should be noted here -- and is true for most other production sites -- that the production rate, $\dot{N}_{+}$, is \emph{not} (necessarily) the annihilation rate, $\dot{N}_{\pm}$.
Most \poss{} are produced at relativistic energies and supposedly propagate in the ISM on Myr timescales until they annihilate, which makes it difficult to judge whether the 511\,keV phenomenon is actually a steady state.
In this context, also the escape of \poss{} from the Galaxy should be considered as they would be lost for current observational techniques\footnote{Coded-mask telescopes, such as the currently best telescope to observe the 511\,keV emission, INTEGRAL/SPI, cannot, or only hardly, observe isotropic emission as they rely on the contrast between emission and no emission to distinguish the instrumental background from the total signal. Shallow gradients in the morphology are also difficult to reliably detect with coded aperture masks \citep{Siegert2022_gammaraytelescopes}}, even if they do annihilate in the IGM, for example.
This means that the \pos{}-production rate in the Milky Way may be larger or smaller than the measured \pos{}-annihilation rate.

The advantage of a nucleosynthesis origin of \poss{} in the Galaxy is that we can directly measure the mass of ejecta, given we have a 3D geometrical model of the Galaxy and the $\gamma$-ray opacity of SN remnants as a function of time.
Because each nucleus can only decay once, and if the possible $\gamma$-ray emission from daughter nuclei can be measured, we obtain a generic luminosity per isotope per solar mass, $L_\odot^i = p_i^{\gamma} \dot{N}_i$, which directly converts to a \pos{}-production rate, given Eq.\,(\ref{eq:positron_production_betaplus}):
\begin{equation}
		\dot{N}^{+}_i = 3.8 \times 10^{43} \frac{p_i}{A_i} \left(\frac{\tau_i}{\mrm{Myr}}\right)^{-1}\left(\frac{M_i}{\mrm{M_{\odot}}}\right)\,\mrm{e^+\,s^{-1},}
	\label{eq:isotope_luminosity}
\end{equation}
where $A_i$ is the nucleon number of isotope $i$.

For example, given the estimate of $2.8\,\mrm{M_\odot}$ of \nuc{Al}{26} in the Milky Way \citep{Siegert2017_PhD,Pleintinger2020_PhD}, and a probability for $\beta^+$-decay of \nuc{Al}{26} of $p_{26} = 82\,\%$, we obtain a production rate\footnote{Detailed modelling then provides the numbers in Tab.\,\ref{tab:source_rates}.} of $3.2 \times 10^{42}\,\mrm{e^+\,s^{-1}}$.
These are at most 10\,\% of the total annihilate rate in the Milky Way, which means that \nuc{Al}{26} alone cannot explain the measured 511\,keV line, unless there was a star burst from which the \poss{} are still annihilating now \citep[e.g.,][]{Prantzos2011_511,Alexis2014_511ISM}.

Considering the shorter-lived isotopes, such as \nuc{Ti}{44}, for example, we have to rely on the only direct measurement of \nuc{Ti}{44} in the Milky Way so far, the SN remnant Cassiopeia A (Cas A), to estimate the production rate.
Assuming all \poss{} from \nuc{Ti}{44} to escape from the remnant, the total number produced is given by $N_{0,44}^+$, and the total production rate by the SN rate in the Milky Way, $R_{\rm SN}$, as
\begin{equation}
	\dot{N}^{+}_{\rm SN,44} = N_{0,44}^+ R_{\rm SN}\mrm{.}
	\label{eq:supernovarate_production}
\end{equation}
With a \nuc{Ti}{44} ejecta mass produced in Cas A to be around $10^{-4}\,\mrm{M_{\odot}}$ \citep[e.g.,][]{Renaud2006_CasA,Martin2008_CasA,Iyudin1997_CasA,Siegert2015_CasA,Weinberger2020_CasA}, and a branching ratio of $0.94$, we get a total number of \poss{} produced of $2.6 \times 10^{51}\,\mrm{e^+}$.
With a ccSN-rate of $R_{\rm SN} = 2/(\mrm{100\,yr})$ \citep[e.g.,][]{Diehl2006_26Al}, the \nuc{Ti}{44} \pos{}-production from ccSNe in the Milky Way is about $1.6 \times 10^{42}\,\mrm{e^+\,s^{-1}}$ -- of the same order of magnitude as \nuc{Al}{26}.
It should be noted that this may even be considered an upper bound of the production rate from \nuc{Ti}{44} because the yield per ccSN may indeed be lower and Cas A exceptional, and because it has been shown in the case of SN1987A that \poss{} are responsible for heating the remnant at early times \citep{Seitenzahl2014_SN87A}.
However, also thermonuclear SNe produce \nuc{Ti}{44}, which enhances the \nuc{Ti}{44} \pos{}-production rate by about a factor of 2.

\paragraph{Type Ia supernovae:}
The case for type Ia supernovae (SNe Ia) in the Galaxy is even more insecure:
While the amount of \nuc{Ni}{56} produced in SNe Ia is about $0.1$--$1.0\,\mrm{M_{\odot}}$ \citep[from Arnett's rule,][]{Arnett1982_SNIa,Stritzinger2006_SNIa}, and the branching ratio for $\beta^+$-decay of the daughter nucleus \nuc{Co}{56} about 20\,\%, the escape fraction of \poss{} from the expanding remnant is very uncertain.
Because \nuc{Ni}{56} and \nuc{Co}{56} have lifetimes of only 8\,d and 111\,d, respectively, it can be expected that most \poss{} do not manage to escape into the ISM but rather annihilate in situ.
\citet{Milne1999_SNIa} showed that the escape of \poss{} is probably on the order of 3\,\%, as estimated from the late lightcurves of a sample of SNe Ia.
The only direct measurement of the \nuc{Ni}{56} mass from a SN Ia is from SN2014J, estimating $0.4$--$0.6\,\mrm{M_{\odot}}$ \citep{Churazov2014_SN2014J,Diehl2015_SN2014J_Co}.
The data suggest that there indeed is a broadened 511\,keV line, but the flux uncertainties make it difficult to reliably estimate a \pos{} escape fraction.
At face value, the average escape fraction $\eta$ from SN2014J during the half-year observation time is $\eta = 0.06^{+0.33}_{-0.06}$ \citep{Siegert2017_PhD}.
Therefore, similar to the \nuc{Ti}{44} case, the \pos{}-production rate from \nuc{Ni}{56} in SNe Ia is
\begin{equation}
	\dot{N}^{+}_{\rm SN,56} = N_{0,56}^+ \eta R_{\rm SNIa}\mrm{.}
	\label{eq:SNIa_production}
\end{equation}
Using the $\gamma$-ray measurements from SN2014J and an estimated SNe Ia rate in Milky Way of $0.25/\mrm{century}$, the \nuc{Ni}{56} \pos{}-production rate is $\left(2_{-2}^{+12}\right) \times 10^{43}\,\mrm{e^+\,s^{-1}}$.
At a face value of $\eta = 0.03$ and a canonical \nuc{Ni}{56} ejecta mass of $0.5\,\mrm{M_{\odot}}$, the production rate would be $5.1 \times 10^{42}\,\mrm{e^+\,s^{-1}}$ -- again of similar magnitude than the previous two nucleosynthetic \pos{} origins.
It is clear that \poss{} from SNe Ia may either explain the complete signal, none of it, or even more than is measured.

\paragraph{Classical novae:}
Classical novae (CNe) have never been detected in MeV $\gamma$-rays, so that a direct estimate of the important ejecta masses has not been possible so far.
In CNe, two contributions may be of interest, either the `511\,keV flash' from very short-lived $\beta^+$-unstable nuclei, such as \nuc{F}{18} (2.64\,h) or \nuc{N}{13} (14.4\,min), or from the longer-lived \nuc{Na}{22} (3.75\,yr).
While the latter is dominantly only produced in ONe novae, the former are expected in all types of CNe.
Models \citep[e.g.,][]{Gomez-Gomar1998_novae,Jose1998_novae,Hernanz2000_novaflash,Jose2011_novae,Hernanz2006_novae,Hernanz2014_nova} suggest a $\gamma$-ray flash at the onset of the CN -- the explosion on the surface of the white dwarf --, several days before the optical emission maximum.
The exact time is unknown, but the flux from the \poss{} annihilating directly on the surface is on the order of $10^{-3}$--$10^{-2}\,\mrm{ph\,cm^{-2}\,s^{-1}}$ at 511\,keV for a distance of 1\,kpc for a duration of about 1\,h.
This would be easily detectable by current $\gamma$-ray telescopes, but has never been observed.
This might, in fact, cast doubts on the strength of the 511\,keV line from the very-short-lived isotopes, and it has been suggested by \citet{Leung2022_novae} that the flux would be heavily suppressed because the initial ejecta velocity could be much smaller than expected, so that the opacity stays at higher values at early times.
With a flux of $10^{-9}$--$10^{-8}\,\mrm{ph\,cm^{-2}\,s^{-1}}$ at 511\,keV, these lines could never be detected with current instrumentation.
In any case, these \poss{} would most probably fall into the in-situ annihilation case unless they also escape.

For \nuc{Na}{22}, we have to rely so far on stellar evolution models and nucleosynthesis calculations to estimate the contribution of CNe to the remaining Positron Puzzle.
While 1/3 of all CNe are considered ONe novae, and the CNe rate in the Milky Way is about $20$--$60\,\mrm{yr^{-1}}$ \citep{Shafter2017_novarate}, the expected contribution of \nuc{Na}{22} \poss{} is $<3 \times 10^{41}\,\mrm{e^+\,s^{-1}}$.
Measurements only find an upper limit on the cumulative flux of \nuc{Na}{22} $\gamma$-rays at 1275\,keV in the Milky Way of $<2 \times 10^{-7}\,\mrm{M_{\odot}}$ per nova, which converts to an upper bound of the \pos{}-production rate of $<5 \times 10^{42}\,\mrm{e^+\,s^{-1}}$ \citep{Siegert2021_BHMnovae} -- about ten times higher than the theoretical expectation.
CNe make at most 10\,\% of the total annihilation rate in the Milky Way if a steady state is assumed, and it is suggested from yield models that their contribution is even less than 1\,\%.
It should be noted here that the contribution of CNe to Galactic \nuc{Al}{26} is still debated, so that the \pos{}-budget for massive stars might be decreased and for CNe increased, but will not change in total.

\paragraph{Other nucleosynthesis origins:}
In the above considerations, we only took into account the standard classes of objects.
Rare but strong events (or similarly weak but frequent events) may contribute significantly to the production of \poss{} in the Galaxy.
There are no direct measurements on which we could base our estimates on, so that here, we only list these objects as possible additions(!) to the total production rate.

Hypernovae, for example, which may be related to pair-instability supernovae, could produce $20$--$40\,\mrm{M_\odot}$ of \nuc{Ni}{56}, as suggested from SN2006gy in NGC1260 \citep{Smith2007_PISN}.
If such an event happened every 1000\,yr, the Milky Way annihilation rate could be sustained by hypernovae alone, even with an escape fraction of 1\,\%.
However, the hypernova interpretation has been contested with other scenarios, for example as interactions with the ISM would strongly decrease the \nuc{Ni}{56} mass \citep[e.g.,][]{Jerkstrand2020_SN2006gy,Fox2015_SN2006gy}.

Another scenario which may explain all the \poss{} in the Galaxy required to meet its annihilation rate is by \citet{Crocker2017_511_91bg}.
In a similar fashion as the superluminous supernovae, a rare type of SN Ia, SN1991bg, may produce large amounts of \nuc{Ti}{44}.
In this scenario, a low-mass He white dwarf merges with a low-mass CO white dwarf to produce a subluminous SN Ia.
Because of the large amounts of He available, the yields of \nuc{Ti}{44} may be as high as $0.03\,\mrm{M_{\odot}}$, $10^2$--$10^4$ times more than expected (and measured) for `normal' ccSNe and SNe Ia.
With a rate of these objects of 1 every 500\,yr, averaging to about 15\,\% of all SNe Ia to be of SN1991bg-type, the total production rate would be on the order of $5 \times 10^{43}\,\mrm{e^+\,s^{-1}}$.
There is indeed evidence that these objects produce Ti as measured from absorption spectra around $420\,\mrm{nm}$ \citep[see also][]{Higdon2009_511}.

\paragraph{Compact objects:}
In compact objects, i.e. in particular neutron stars (NSs) and BHs, pair production by photon-photon interactions is probably the dominant \pos{} production channel.
Despite the fact that the production cross section of photon-photon-pair-production is calculated as
\begin{equation}
	\sigma_{pair} = \frac{3\sigma_T}{16}(1-\beta^2)\left[(3-\beta^2)\ln\left(\frac{1+\beta}{1-\beta}\right) - 2\beta(2-\beta)^2\right]\mrm{,}
\end{equation}
which has a maximum at $\beta = 0.7$ for $\sigma_{\rm pair}^{\rm max} \approx 1.7 \times 10^{-25}\,\mrm{cm^2}$, i.e. on the same order of magnitude as the Thomson cross section, photon-photon-pair-production has never been unambiguously observed in laboratory experiments.
Due to the high densities of high-energy photons needed, this process is only efficient when the optical depth $\tau_{\gamma\gamma} = n_{\gamma}\sigma_{\rm pair}x$, with $x$ being the physical size of the emitting objects, is larger than unity.
This can be expressed in terms of the so-called compactness parameter $l$, as
\begin{equation}
	l \equiv \frac{L\sigma_T}{rm_ec^3} = 2\pi \left(\frac{m_p}{m_e}\right) \left(\frac{L}{L_E}\right) \left(\frac{r_S}{r}\right)\mrm{,}
	\label{eq:compactness}
\end{equation}
where $L_E$ is the Eddington luminosity and $r_S$ is the Schwarzschild radius of the source, both being a function of the object's mass \citep{Lightman1987_compactness}.
For efficient pair production, $l \gtrsim 10$, which means that for small objects and/or large luminosities, \emph{pair production is inevitable}.
Such compactness values are naturally possible for BHs, NSs, and to a smaller extent also WDs.

X-ray binaries (XRBs), and in particular low-mass XRBs (LMXRBs) were discussed in terms of the `Positron Puzzle' shortly after the launch of INTEGRAL \citep{Prantzos2004_ns}.
They raised even more attention because their spatial distribution appeared similar to that of the 511\,keV image with an enhancement towards negative longitudes \citep{Weidenspointner2008_511}.
It turned out that the INTEGRAL/SPI measurements at this point in time were not representative of the full Galaxy so that a claim considering an XRB-origin due to the similar spatial distribution is flawed.
Nevertheless are XRBs promising \pos{}-producers as the annihilation of the pair-plasma that is produced in these objects has supposedly been measured in three different objects, the `Great Annihilator' \citep[1E1740.7-2942,][]{Bouchet1991_mq}, Nova Musca \citep[GRS1124-68,][]{Goldwurm1992_511,Sunyaev1991_mq}, and V404 Cygni \citep[GS2023+338,][]{Siegert2016_V404}.
All of these measurements are contentious because independent analyses could not confirm the, most of time, weak signals.

The models for pair production in XRBs \citep[e.g.,][]{Beloborodov1999_511}, their annihilation as plasma \citep[e.g.,][]{Svensson1983_TPA}, and possible escape into the ISM \citep[e.g.,][]{Beloborodov1999_511} result in order of magnitude estimates for the total population of compact objects in the Galaxy.
Given the measurements of V404 Cygni, for example, we can estimate a \pos{}-production rate of $(1.1 \pm 0.3) \times 10^{42}\,\mrm{e^+\,s^{-1}}$.
Note that even if the pair-plasma annihilation feature in the spectra are doubted, the MeV spectrum extends significantly above 511\,keV, so that pair production is inevitable \citep{Svensson1987_TPA,Maciolek-Niedzwiecki1995_xrb511}.
In the case of V404 Cygni with a luminosity of 20\,\% Eddington \citep{Rodriguez2015_v404}, the photon-photon absorption has to take place within $\approx 10 r_S$.
This leads to an efficient absorption of $\gamma$-rays as they can interact with the more numerous X-rays, which leads to a suppression of photon energies $\gg m_ec^2$.
Estimates of the GeV flux of $<10^{-6}\,\mrm{ph\,cm^{-2}\,s^{-1}}$ corroborate this scenario \citep{Siegert2016_V404,Harvey2021_v404}.

Assuming a power-law-shaped seed spectrum \citep{Beloborodov1999_511}, the luminosity that potentially contributes to the \pos{}-production in XRBs can be estimated.
Assuming further that the $\gamma$-ray measurements in V404 Cygni are indeed pair-plasma-annihilations, the escape fraction of \poss{} can be estimated to be $\eta_{\gamma\gamma} = (60 \pm 20)\,\%$.
The theoretical value should be at most 20\,\%; other analyses of the same source would suggest an escape close to 100\,\% \citep{Roques2015_V404,Jourdain2017_V404}.
These values are all of the same order of magnitude, and a canonical production rate per XRB in outburst in the Milky Way can be given as $\dot{N}^+_{\rm \gamma\gamma,XRB} = 10^{42}\,\mrm{e^+\,s^{-1}}$ -- corroborated by measurements.

To estimate the total contribution of microquasars and other XRBs to the \pos{}-production in the Galaxy, one needs to estimate the duty cycle of such sources, and the total number of objects from population synthesis calculations \citep{Guessoum2006_MQ511}.
The duty cycle $\epsilon_{\rm XRB}$ may be on the order of $10^{-3}$--$10^{-2}$, since per year, about one XRB is found in outburst for about 1--10\,d.
The total number of XRBs in the Galaxy may be in the range of $N_{\rm XRB} \approx 10^4$ as suggested by several works \citep[e.g.,][]{Romani1992_LMXRB,Portegies_Zwart1997_LMXRB,Bandyopadhyay2009_511}.
The total number of \poss{} from the population of XRBs in the Milky Way is then
\begin{equation}
	\dot{N}^{+}_{\gamma\gamma} = N_{\rm XRB} \epsilon_{\rm XRB} \dot{N}^+_{\rm \gamma\gamma,XRB} = (3 \pm 3) \times 10^{43}\,\mrm{e^{+}\,s^{-1}.}
	\label{eq:XRB_production}
\end{equation}

\noindent Again, at face value, and motivated by observations, XRBs may explain the entire annihilation rate in the Galaxy if a steady state is assumed.
However because their escape is uncertain, the seed spectrum is uncertain, the duty cycle is uncertain, and the number of objects in the Galaxy is uncertain, the population of XRBs may also contribute insignificant amounts of \poss{} for the Positron Puzzle.

\paragraph{Pulsars:}
The contribution from pulsars, i.e. pair production in magnetic fields, can only be estimated theoretically, or by considering the measurements of cosmic-ray \poss{} in the Solar System.
Based on geometrical models, the \pos{}-production rate by pulsars can be estimated as
\begin{equation}
	\dot{N}^+_{B\gamma} \approx 2.8 \times 10^{37} \left(\frac{B}{\mrm{10^{12}\,G}}\right)^{10/7} \left(\frac{P}{\mrm{s}}\right)^{-8/21}\,\mrm{e^{+}\,s^{-1},}
	\label{eq:pulsar_production}
\end{equation}
where $B$ is the dipole magnetic field strength of the pulsar and $P$ its rotation period \citep{Cheng1986_pulsars,Zhang1997_pulsars}.
Given the population of pulsars in the Milky Way, including normal ones, millisecond pulsars, and magnetars\footnote{For millisecond pulsars and magnetars, the dipole assumptions may not be valid anymore.}, a total production rate of $\dot{N}^+_{\rm Pulsars} \approx 5 \times 10^{42}\,\mrm{e^+\,s^{-1}}$ can be estimated.

\begin{figure}[t]
	\centering
	\includegraphics[width=\columnwidth,trim=0.0cm 0.3cm 2.0cm 1.0cm,clip]{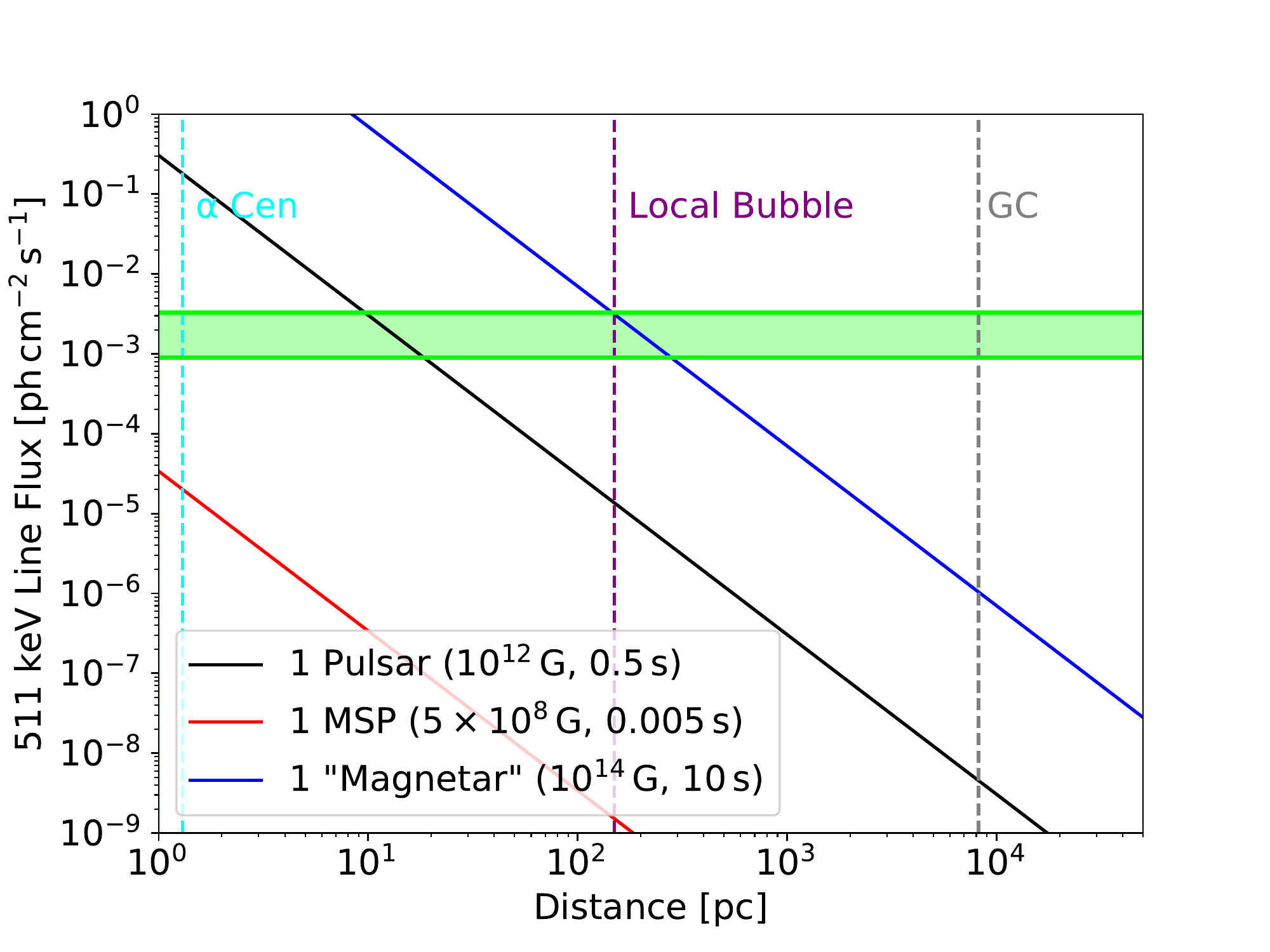}
	\caption{The `one neutron star solution' of the Positron Puzzle: The flux of the Galactic 511\,keV line (green band) placed at a distance of 100-200\,pc can be explained by the production rate of a single magnetar (blue line, magnetic field $10^{14}$\,G). Smaller fluxes, such as the $8 \times 10^{-5}\,\mrm{ph\,cm^{-2}\,s^{-1}}$ from the \emph{direction} of the Galactic centre, may be explained by a single pulsar with a magnetic field of $\sim 10^{12}$\,G.}
	\label{fig:one_neutron_star}
\end{figure}

Interestingly, Eq.\,(\ref{eq:pulsar_production}) can be used to estimate the distance a single pulsar must have to explain the entire 511\,keV flux from the direction of the Galactic bulge:
In this `one neutron star' solution, the 511\,keV flux of $(1$--$3)\times 10^{-3}\,\mrm{ph\,cm^{-2}\,s^{-1}}$ from the entire Milky Way is used and placed at a distance $d$ away from the Solar System.
Clearly, this estimate suggests that all \poss{} slow down in a way that the 511\,keV morphology is recovered, which may not be the case, but the scenario is easily scalable.
Assuming that the pulsar \pos{}-production rate and observed flux are a steady state, the resulting plane of $F_{511}$-$d$ can be found in Fig.\,\ref{fig:one_neutron_star}.
Given typical numbers for normal, millisecond pulsars (MSPs), and highly magnetised NSs (magnetars), the entire measured 511\,keV line flux can be explained by a single magnetar at a distance of 100--200\,pc.
Smaller fluxes may still be explained by regular pulsars at such a distance.

It is intriguing that the high-energy \pos{}-excess from cosmic-ray measurements with AMS-02 also suggests a nearby ($\lesssim 250$\,pc) and young ($\approx 0.1$\,Myr) pulsar \citep{Aharonian1995_pospulsar,Faherty2007_geminga}.
Indeed, including the Geminga pulsar significantly improves the description of the AMS-02 data \citep{Yin2013_pulsarAMS,DiMauro2014_pospulsar}.

The question emerges: `why not relating the two problems?'
The canonical answer to this question is that `the kinetic energies of the \poss{} would be too large.'
As I will argue in Sec.\,\ref{sec:propagation}, such a statement may be flawed, first in terms of astrophysics, and second in terms of analysis.
This dogma reappears also in the case of general cosmic-ray \pos{}-production as well in the case of DM.

\begin{figure*}[t]
	\centering
	\includegraphics[width=\textwidth]{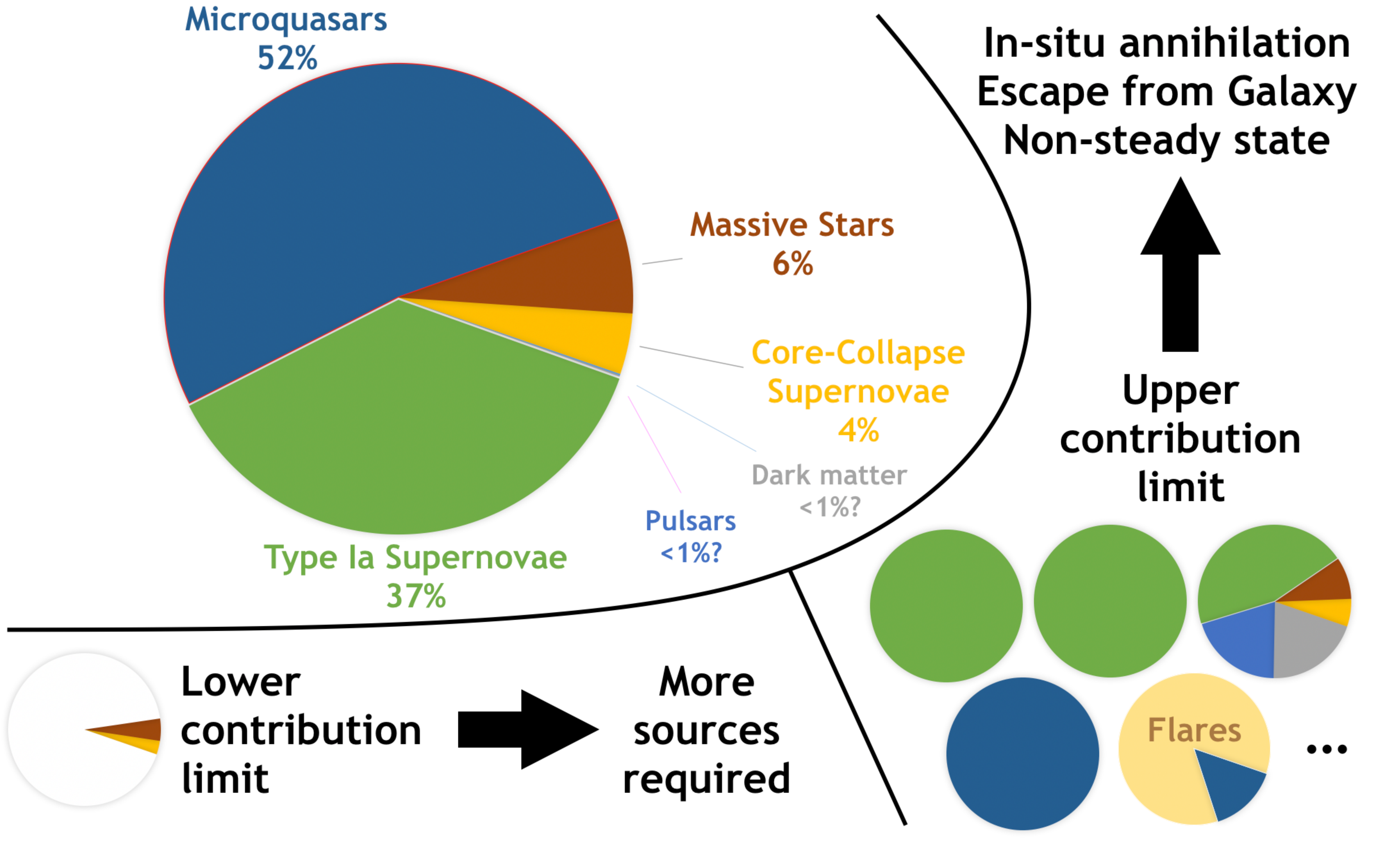}
	\caption{Positron production rate in the Milky Way. \textbf{\emph{Top left:}} About 105\,\% of the \poss{} \emph{seen} to annihilate are composed of measured $\beta^+$-emitters (SNe Ia, massive stars, ccSNe, in total $\approx 47\,\%$) and microquasars in outburst ($\approx 52\,\%$). DM, pulsars, cosmic-ray secondaries, and in general \poss{} sources with Lorentz factors $\gamma \gtrsim 100$, might not contribute at all to the 511\,keV line. \textbf{\emph{Bottom left and right:}} Taking into account the measurement uncertainties in both production and annihilation, as well as the uncertainties in population sizes and additional model variants, may either result in at most 10\,\% production vs. annihilation so that more sources would be required, or more than 500\,\% production vs. annihilation which might indicate a non-steady-state scenario, significant escape from the Galaxy, or dominant in-situ annihilation \citep{Siegert2017_PhD}.}
	\label{fig:positron_cake_diagrams}
\end{figure*}

\paragraph{Supermassive black holes -- the case of Sgr A*:}
Similar to stellar mass black holes, supermassive black holes (SMBHs) with masses of $10^6$ to $10^9\,\mrm{M_{\odot}}$ could also produce \ele{}-\pos{}-pairs in the vicinity of their event horizons, near the hot inner accretion disk, above in their coronae, or in relativistic jets.
For the SMBH in the centre of the Milky Way, Sgr A* \citep[e.g.][]{Genzel2003_Sgra,Ghez2004_SgrA}, the accretion history is the most important factor in the context of the `Positron Puzzle'.
In the quiescent state it is now, its accretion rate is on the order of $10^{-8}$--$10^{-6}\,\mrm{M_{\odot}\,yr^{-1}}$ \citep{Baganoff2003_SgrAstar}.
If Sgr A* had phases of higher activity, an accretion rate of up to $10^{-4}\,\mrm{M_{\odot}\,yr^{-1}}$ might have been possible, which would have increased the inner accretion disk temperature up to $10^{11}$\,K \citep{Totani2006_accretion}.
Such a large accretion rate may be due to tidal disruption events of low-mass stars \citep{Cheng2007_SgrAstardiffuse}, massive star winds \citep{Quataert2004_SgrAstardisrupt}, or from the ionised halo surrounding the SMBH \citep{Totani2006_accretion}.
In either case, this may be enough to produce thermal MeV $\gamma$-rays, which in turn would create \ele{}-\pos{}-pairs, but on a much shorter timescale than the annihilation is expected to happen.

The model by \citet{Totani2006_accretion} suggests that Sgr A* alone could provide a 511\,keV line flux on the order of $10^{-3}\,\mrm{ph\,cm^{-2}\,s^{-1}}$ if the accretion rate about 300\,yr ago was $10^3$--$10^4$ times higher than now.
Another possibility could be periodic tidal disruption events with a timescale of 100\,kyr \citep{Cheng2007_SgrAstardiffuse}, or episodic interruptions of \pos{} production in Sgr A* due to quenching from nearby SN explosions \citep{Totani2006_accretion}.
Steady production of \poss{} due to proton-proton interactions in the vicinity of Sgr A* would then occur on shorter timescales than in the ISM, for example.
The \poss{} might then diffuse outwards and fill up the bulge, but which has been shown to come with energetic constraints \citep{Panther2018_nuclear_outflow}.

In any case, the measured point-like source in the centre of the Galaxy \citep{Skinner2014_511,Siegert2016_511} may be due to \poss{} from Sgr A*, but the narrow 511\,keV line instead of the expected pair-plasma feature for BHs would point to the enhanced gas density in the CMZ rather than the SMBH itself.

\paragraph{Cosmic ray secondaries and $\gamma$-ray albedo of the Solar System:}
Cosmic rays that have been accelerated by SN remnants, for example, reaching up to $\approx 10^{15}$\,eV, in addition to ultra-high-energy cosmic rays up to $10^{21}$\,eV, can lead to secondary \poss{} as a result of high-energy collision.
\citet{Porter2008_CR} estimated the secondary \pos{}-production from cosmic rays by propagation models, taking into account energy losses, diffusion, and escape, to be between $(1$--$2) \times 10^{42}\,\mrm{e^+\,s^{-1}}$.
Again, the contribution is on the order of a few percent of the total measured annihilation rate, but this time with large kinetic energies, peaking around $30$--$40$\,MeV and extending to several GeV--TeV \citep{Murphy1987_pionpositrons}.
Such \poss{} may again be `too fast' to explain the Galactic 511\,keV signal for which eV-scale particles are required (see Sec.\,\ref{sec:low-energy-propagation}).

In addition to secondary \pos{}-production of cosmic rays with interstellar gas and molecular clouds, primary cosmic rays also interact with solid bodies in the Solar System.
\citet{Moskalenko2007_GRalbedoMoon} and \citet{Moskalenko2008_GRalbedoSS} calculated the $\gamma$-ray spectrum of cosmic rays interacting with small Solar System bodies as well as the Earth Moon.
The 511\,keV line from rapid energy losses of secondary \poss{} in the rock material followed by direct annihilation is on the order of $2 \times 10^{-4}\,\mrm{ph\,cm^{-2}\,s^{-1}}$ distributed along the ecliptic.
This emission would in fact be time variable as the asteroids and trojans move significantly with respect to the Earth orbit around the Sun, so that different hot spots may emerge at different times (Siegert 2023, in prep.; see also Sec.\,\ref{sec:foreground}).
This might also be the reason why this emission component has so far not been detected.
The Moon would show a 511\,keV line on the order of $2 \times 10^{-5}\,\mrm{ph\,cm^{-2}\,s^{-1}}$, and would make a significant foreground component if observed with large-field-of-view instruments.
The quiet Sun, Earth itself and planets are also plausible 511\,keV contributors from cosmic-ray bombardment.

\paragraph{Dark matter:}
Dark matter density profiles to explain the observed phenomena of flat rotation curves, structure formation, and high mass-to-light ratios in dwarf galaxies, are reminiscent of the 511\,keV morphology, without the disk.
It can be shown that an NFW profile squared, as suggested from the co-annihilation of DM particles, can replace the three bulge components to describe the $\gamma$-ray data statistically \citep{Skinner2014_511}.
There are, however, more astrophysically relevant templates that fit the data better \citep{Siegert2021_BDHanalysis}, such as a combination of a (smoothed) infrared bulge \citep{Freudenreich1998_BoxyBulge_COBE} plus a (smoothed) nuclear stellar cluster \citep{Launhardt2002_NB}.
If DM particles are on the GeV--TeV mass scale, multiple \eles{} and \poss{} are emitted from their annihilation, resulting in a strong contribution to the IC spectrum in the Milky Way at MeV energies.
These DM annihilation products have also been invoked to explain the cosmic-ray \pos{}-excess as an alternative to the nearby pulsar scenario \citep[see review by, e.g.,][]{Serpico2011_positronexcess}.

Associating the 511\,keV line in the Milky Way to DM products is difficult because, as shown above, the contribution of many other sources would produce an irreducible fore- and background.
Other galaxies, and in particular dwarf galaxies, may provide better targets to distinguish 511\,keV models.
\citet{Siegert2016_dsph} found a hint at 511\,keV from the dwarf galaxy Reticulum II (Ret II) -- contemporary to a hint at GeV energies \citep{Geringer-Sameth2015_RetII}.
With increased exposure and refined data analysis, it was shown that this signal was a fluke.
It was further argued by \citet{Siegert2016_dsph} that the DM contribution to the 511\,keV signal in the Milky Way, based on the serendipitous signal from Ret II, would be at most 1\,\%.
Because the signal in Ret II may in fact be smaller than what has been measured before, and assuming no astrophysical background in Ret II to the 511\,keV signal, the previous statement only holds if the DM \poss{} cool down quickly and annihilate fast without escaping to the IGM.
Since the only loss would be due to IC scattering, the escape is probably inevitable, but would also show a strong contribution again at the MeV continuum.

If the DM particles, instead, are of MeV scale, the behaviours would be similar to that of nucleosynthesis and pair production origins.
As will be discussed in Sec.\,\ref{sec:propagation}, the propagation may either be ballistic (collisional) for which large distances could be travelled within a typical \pos{} lifetime of 0.1--10\,Myr in the ISM, or advective, following the Galactic gas flow with diffusion length scales on the order of 1\,pc due to the small gyro-radii.
MeV \poss{} may either emerge from particle DM or PBHs in the mass range of $10^{14}$--$10^{17}$\,g.

It is difficult to estimate relevant parameters, such as DM annihilation cross sections, from the measurements of the Milky Way alone.
Attempts that claim to be `conservative' and overplot the maximum possible spectrum at 511\,keV from a population of DM particles in the Milky Way are not useful:
Gamma-ray data analyses rely on model expectations versus measured count rates -- a model comparison that uses already-extracted fluxes, i.e. values that already assumed a model (e.g. a power-law per energy bin), is flawed and can lead to much \emph{tighter} limits than the anticipated `conservative' approach.
Instead, all models that are required to explain the data should be included, so that their intrinsic degeneracy given the measurement methods, are taken care of.
The uncertainties per model component can and will be much larger than the sum of all components because of the covariance between components.
This means that overplotting DM models on total spectra of the Milky Way in the keV--TeV range is wrong and upper limits from such attempts are not meaningful.

In addition to DM annihilation to pairs, also their internal bremsstrahlung should be taken into account, their IC scattering off the interstellar radiation field and the CMB, the annihilation in flight of \poss{}, and the final annihilation resulting in the 511\,keV line and ortho-Ps spectrum.
The renewed analyses of Ret II is doing that (except for IC) and determines upper bounds on DM decay and annihilation cross sections as well as PBH mass limits.
The velocity averaged cross section has an upper bound of $\langle \sigma v \rangle \lesssim 5 \times 10^{-28} \left(m_{\rm DM}/\mrm{MeV}\right)^{2.5}\,\mrm{cm^3\,s^{-1}}$ and excludes PBH masses to make the gross of DM of $\lesssim 8 \times 10^{15}$\,g \citep{Siegert2021_RetII}.
The particle DM limits are among the strongest compared to current literature \citep[see also][for more limits]{Laha2020_PMBHDM,Slatyer2016_CMB_DM,Ng2019_DM_NuSTAR_M31}.
The strongest PBH evaporation limit is derived from the Cosmic Gamma-ray Background (CGB) including the redshift-integrated 511\,keV contribution as well as the isotropic DM halo component of the Milky Way by \citet{Iguaz2021_CGBPBH}, and from the DM halo of the Milky Way itself, taking into account any possible foreground to obtain a DM-only spectrum by \citet{Berteaud2022_SPI_PBH}.
From BH evaporation, PBH masses of $\lesssim 4 \times 10^{17}$\,g are excluded to account for 100\,\% of the DM mass in the Milky Way.
To exclude higher masses, the keV emission in the Milky Way must be studied as also fewer \poss{} will be produced.

\paragraph{Stellar flares:}
From RHESSI, SMM, Fermi/GBM, and INTEGRAL measurements, for example, we know that solar flares can lead to a strong 511\,keV line and ortho-Ps continuum due to particle acceleration and subsequent production of pions and their decay.
Whether the line is ubiquitous for all flare strengths and for all stars is a plausible, yet unproven assumption.
From Kepler and  TESS measurements \citep[e.g.,][]{Shibayama2013_flares}, it is known that also other stars flare in a similar fashion as the Sun, with a flare-frequency-energy-distribution of $\frac{dN}{dEdt} \approx 10^{-29}\left(\frac{E}{10^{30}\,\mrm{erg}}\right)^{-1.8}\,\mrm{erg^{-1}\,yr^{-1}}$.
From the rare X-class flares with energies above $10^{31}$\,erg, RHESSI could measure high-resolution spectra in the MeV range, finding also nuclear excitation lines in addition to the strong neutron capture line at 2.2\,MeV and the \pos{}-annihilation line at 511\,keV \citep{Lin2002_RHESSI}.
\citet{Murphy2005_posiloss}, \citet{Murphy2014_511flares}, and \citet{Share2004_solarflare511} studied the annihilation of \poss{} in the solar atmosphere during flares in great detail, including line shapes and Ps-fractions as functions of temperature, density, and ionisation state:
For low-energy ions, \nuc{He}{3} and $\alpha$-particle reactions are the dominant contributors to $\beta^+$-unstable isotopes during a flare, resulting in large abundances of \nuc{C}{11}, \nuc{N}{13}, \nuc{O}{15}, \nuc{F}{17}, and \nuc{F}{18}, depending on the initial \nuc{He}{3}/$\alpha$ ratio and the ion spectral index.
As soon as the energy per ion reaches beyond $\approx 300\,\mrm{MeV\,nucleon^{-1}}$, $\pi^+$-production via proton-reactions is the only significant contribution to the \pos{}-content.
The slowing-down timescale from such \poss{} in the solar atmosphere is on the order of 20\,s \citep{Murphy2014_511flares} so that a significant amount of \poss{} annihilate directly in the atmosphere.
Escape into the interplanetary medium may also be possible due to this timescale.

X-class solar flares are rare, but the only ones in which the 511\,keV line has been detected, yet.
Assuming that all flares produce \poss{} which annihilate either directly in the stellar atmosphere or escape the star and annihilate in their vicinity, \citet{Bisnovatyi-Kogan2017_511} estimated from a benchmark solar flare that 50--100\,\% of the 511\,keV line in the bulge may indeed be due to intermittently flaring stars.
The model can easily be extended to include the Galactic disk.
A relation between the 511\,keV line flux and flare energy would refine this scenario as it is not clear how many \poss{} are really produced per flare and how many annihilate in situ.
Such a scenario would immediately reduce the need for many different source types because the estimated and measured number for nucleosynthesis \poss{} would now only need to explain a small fraction of the total flux.
It would \emph{also} reduce the need for far propagation (Sec.\,\ref{sec:propagation}) because, as stated in Sec.\,\ref{sec:imaging}, the stellar population already fits the data quite well.
See Fig.\,\ref{fig:positron_cake_diagrams} for an overview of the production rates of different models assumptions vs. the measured annihilation rate in the Milky Way.

A smoking gun evidence for a stellar flare scenario would be the detection of 511\,keV lines from globular clusters that correlate with the number of stars in each cluster.
Alternative scenarios in globular clusters are also possible, but they might show a different correlation, e.g. with the GeV flux from MSPs \citep{Bartels2018_binaries511}.

\section{Positron propagation}\label{sec:propagation}
\subsection{High-energy positrons}\label{sec:high-energy-propagation}
The propagation of \poss{} in the ISM is not well understood.
From AMS-02, PAMELA, and other particle detectors around Earth, we know that at least high-energy \poss{}, with Lorentz-factors of $\gtrsim 1000$ reach Earth.
Lower energy \poss{} are, similar to the case of other cosmic-ray species, heavily affected by the solar modulation potential.
There appears to be an excess of cosmic-ray \poss{} above the expectations from secondary production of hadronic cosmic rays with the ISM that dominates the spectrum above $\approx 20$\,GeV \citep{Aguilar2013_AMS,Aguilar2019_AMS02_positrons}.
This component has been interpreted in terms of a nearby ($\lesssim 250$\,pc) pulsar that manages to accelerate \poss{} to Lorentz factors of $\gtrsim 10^6$ \citep[e.g.,][]{DiMauro2014_pospulsar,Yin2013_pulsarAMS}, or in terms of heavy ($\approx 1.2$\,TeV) DM particles, co-annihilating into pairs \citep[e.g.,][]{Kopp2013_AMSdarkmatter}.
While the first scenario may have been excluded by HAWC measurements \citep{Abeysekara2017_AMSpulsarHAWC}, the latter may be corroborated by ongoing measurements of AMS-02 within the next years.
The fundamental problem in all of the scenarios is the assumption that cosmic-ray secondary production and \pos{}-propagation in the Galaxy is understood.

\citet{Cowsik2014_CRpropagation} showed that the positron fraction $R_{e^+} = \mrm{e^+/(e^++e^-)}$, as measured by AMS-02, can be explained without additional sources.
In their Nested Leaky-Box model, acceleration of cosmic rays happens in many sources across the Galaxy and \pos{}-propagation follows the established works of \citet{Moskalenko1998_CRpos}, resulting in a source term with a power-law spectrum and spatial distribution according to the hydrogen density in the Galaxy.
The propagation can then be described as spatial diffusion with radiative energy losses, and the leakage from the Galaxy, so that \poss{} may in fact escape on a characteristic time scale.

In summary, for highly relativistic \poss{}, even though we can measure directly the cosmic-ray spectrum at the position of the Solar System, it is not established, how \poss{} propagate in the ISM.
An inaccurate assumption on the high-energy propagation led to conjectures of additional sources, but which may not be required at all if the premise is wrong.

\subsection{Low-energy positrons}\label{sec:low-energy-propagation}
Now, in terms of low-energy \poss{}, either having Lorentz-factors of 1--3 in the case of nucleosynthesis \poss{} or a few 10s to 100s in the case of pulsars, the gross of secondary \poss{} from cosmic-ray interactions in the ISM \citep{Murphy1987_pionpositrons}, pair-plasma ejections from XRBs, light DM particle annihilations, or Hawking radiation, we have no direct information on these particles.
Therefore, we have to rely on theoretical assumptions as to how long and how far \poss{} propagate in the ISM, how many escape the Galaxy, how many annihilate in flight, and how many thermalise with the ISM.

\paragraph{Thermalisation:}
The annihilation rates for cooled \poss{} ($\lesssim 10^7$\,K; $\lesssim 1$\,keV) in interstellar gas as a function of temperature are well understood \citep{Guessoum2005_511}, so that annihilation after thermalisation occurs inevitably after a characteristic timescale of about 1\,Myr for ISM conditions of $T = 8000$\,K and $n_{\rm H} = 1\,\mrm{cm^{-3}}$.
\citet{Panther2018_alkali511} showed that including the much less abundant alkali metals in the ISM reduces this timescale to $0.1$\,Myr.
Propagation after thermalisation is expected to follow the flow of the gas in which the \poss{} are thermalised, i.e. they will be advected rather than transported diffusively, and annihilate according to the conditions in the ISM.
This is then either by charge exchange with hydrogen (only possible for kinetic energies $>6.8$\,eV, i.e. in gas with temperatures not much cooler than $5 \times 10^4$\,K, but also not hotter than $13.6$\,eV ($10^5$\,K) as all hydrogen will be ionised) or by radiative recombination with free \eles{}.
The gas and \ele{} densities will determine the annihilation rate, given the \pos{} density distribution and motion.

\paragraph{Collisional transport:}
The processes between the injection of mildly relativistic \poss{} and the formation of Ps and final decay have been investigated in many studies \citep[e.g.,][]{Prantzos2006_511,Jean2009_511ISM,Martin2012_511,Alexis2014_511ISM,Higdon2009_511,Rothschild2009_511propagation,Siegert2021_BDHanalysis}.
In collisional transport, \poss{} propagate almost ballistically and are guided by the large-scale magnetic field in the Galaxy.
Through pitch-angle scatterings in a turbulent magnetic field, the generic stopping distance for \poss{} in typical ISM conditions is only reduced by 25\,\% \citep{Jean2009_511ISM}, resulting in an effective distance of $10 \left(\frac{n}{\mrm{cm^{-3}}}\right)^{-1}$\,kpc.
At MeV energies, \pos{} energy losses are mainly due to ionisation and excitation of atoms, i.e. in general Coulomb interactions.
Inverse Compton scattering, bremsstrahlung and synchrotron losses are only important for higher energies ($\approx$\,GeV) or when the gas density is small ($\lesssim 10^{-3}\,\mrm{cm^{-3}}$), such as in the Galactic halo or in superbubbles.
On the one hand, this length scale is larger than the typical phases of the warm ISM phase in which most \poss{} presumably annihilate as inferred from spectral analyses \citep{Jean2006_511}.
On the other hand, \citet{Churazov2011_511} showed that the 511\,keV line profile is also consistent with annihilation in a cooling ISM, or within gas phases of temperatures around $100$\,K with only small ionisation fractions on the percent level.
Both of these solutions appear realistic as the annihilation rate increases significantly when \poss{} approach denser regions, but the mixing from the hot ionised gas with HI regions or molecular clouds, for example at the edges of superbubbles, allows also warmer regions and a temperature and thus ionisation gradient \citep{Krause2013_superbubbles}.
These ambiguous solutions may also imply that the assumed turbulence power spectrum has another shape, or that \pos{}-propagation occurs through collisionless transport.

\paragraph{Collisionless transport:}
In collisionless transport, \poss{} propagate according to pitch-angle scattering which may be described by an angular diffusion coefficient, and (re-)acceleration in a turbulent plasma which is governed by a momentum diffusion coefficient.
The basis of this approach is that the gyroradii of $\gamma=1$--$20$ \poss{} in multiphase ISM conditions ($B \approx 1$--$10\,\mrm{\mu G}$), are on the order of $10^8$--$10^9$\,cm, much smaller than the ballistic length scale, so that the exact path of the \pos{} can be averaged over the drift length scale.
The cooling in collisionless transport then happens via particle interactions with magnetohydrodynamic (MHD) waves.
In addition to \pos{}-wave interactions, adiabatic losses, for example in Galactic `winds' or expanding superbubbles, may be significant in this regime.

It is unknown how \poss{} propagate in the ISM:
Different approaches that test the collisional or collisionless transport, either using Monte Carlo simulations \citep{Alexis2014_511ISM} or state-of-the-art cosmic-ray propagation codes \citep{Martin2012_511}, fall short in explaining the 511\,keV morphology as measured with INTEGRAL/SPI.
For example, \citet{Alexis2014_511ISM} expect that the spiral arm structure of the Milky Way would also be seen at 511\,keV.
Given the latest measurements, this is not detected.

\subsubsection{Cosmic-ray transport equation}\label{sec:CRtransport}
In general, the distribution of \poss{} as a function of time, position, and momentum, can be described as the solution of an appropriate Fokker-Planck-equation with \emph{drift} (re-acceleration, convection, advection, energy losses), \emph{diffusion} (coefficients due to transport mode and conditions), \emph{source} (see Sec.\,\ref{sec:sources}), and \emph{sink} (annihilation rates, cross sections, escape, leakage timescales) terms.
Such a treatment requires realistic 3D-distributions of \pos{} sources that are parametrised by their production rate and injection spectrum.
As the $\beta$-decay spectrum is unique for different isotopes, the pair-plasma spectrum nearly thermal, pulsar \poss{} accelerated resulting in a power-law, and cosmic-ray secondary \poss{} peaking around 30\,MeV, one cannot assume one generic spectrum for the injection in the Galaxy -- and certainly no mono-energetic \poss{}.
Furthermore, the gas (target) distributions cannot just be generic, smooth, models because the annihilation rates depend strongly on the varying temperature, ionisation state, target density, and current \pos{}-density.
The energy losses also change as a function of the same parameters, so that an average gas component in such a model will never show the structure of where and at what temperature \poss{} really annihilate.
Diffusion will also change as a function of Galactic component:
For example could the Galactic wind from superbubble blowouts \citep{Krause2021_26Alchimneys} be a region of either ballistic or advective transport up to 10\,kpc into the halo.
Above such a height, turbulence might set in again so that stronger diffusion could be expected.
Similar arguments may be true for the bulge and disk as discussed in \citet{Higdon2009_511}.
An important, albeit often neglected, term in the diffusion-convection equation is the destruction of \poss{}:
It is typically assumed that \poss{} are removed from the scheme if their kinetic energy is below a threshold of, say, 100\,eV \citep{Martin2012_511}.
This assumption may be flawed since the cooling down time scale for MeV \poss{} is on the order of Myr whereas the annihilation timescale after thermalisation might be similar.
The timescales depend on the exact conditions of the ISM at the time, but the cooling down time is generally larger than the annihilation time after thermalisation \citep{Guessoum2005_511,Jean2009_511ISM}.
Nevertheless, in a time of 0.1--1\,Myr, interstellar gas, especially in low-density regions, may be propagating 10--100\,pc and therefore significantly changing the annihilation morphology.

\subsubsection{Annihilation in flight}\label{sec:AiF}
\citet{Martin2012_511} also included the process of \pos{}-annihilation `in flight', i.e. the direct annihilation of \poss{} with \eles{} without forming Ps first, that also removes \poss{} from the scheme.
This process is only governed by the annihilation cross section as a function of energy, given the densities of the interacting particles.
There appears to be a consensus that annihilation in flight is negligible for MeV \poss{} because of two reasons:
1) The probability that this process is happening at higher energies is very low, and 2) \citet{Beacom2006_511} and \citet{Sizun2006_511} claim that the annihilation in flight of \poss{} must not happen at energies above several MeV due to spectral constraints.
I will argue here that both these reasons should be considered only on the premise of a steady state between production and annihilation \emph{inside} the Galaxy, and that the model comparison to the available data was inadequately performed.

\begin{figure*}[t]
	\centering
	\includegraphics[width=0.49\textwidth]{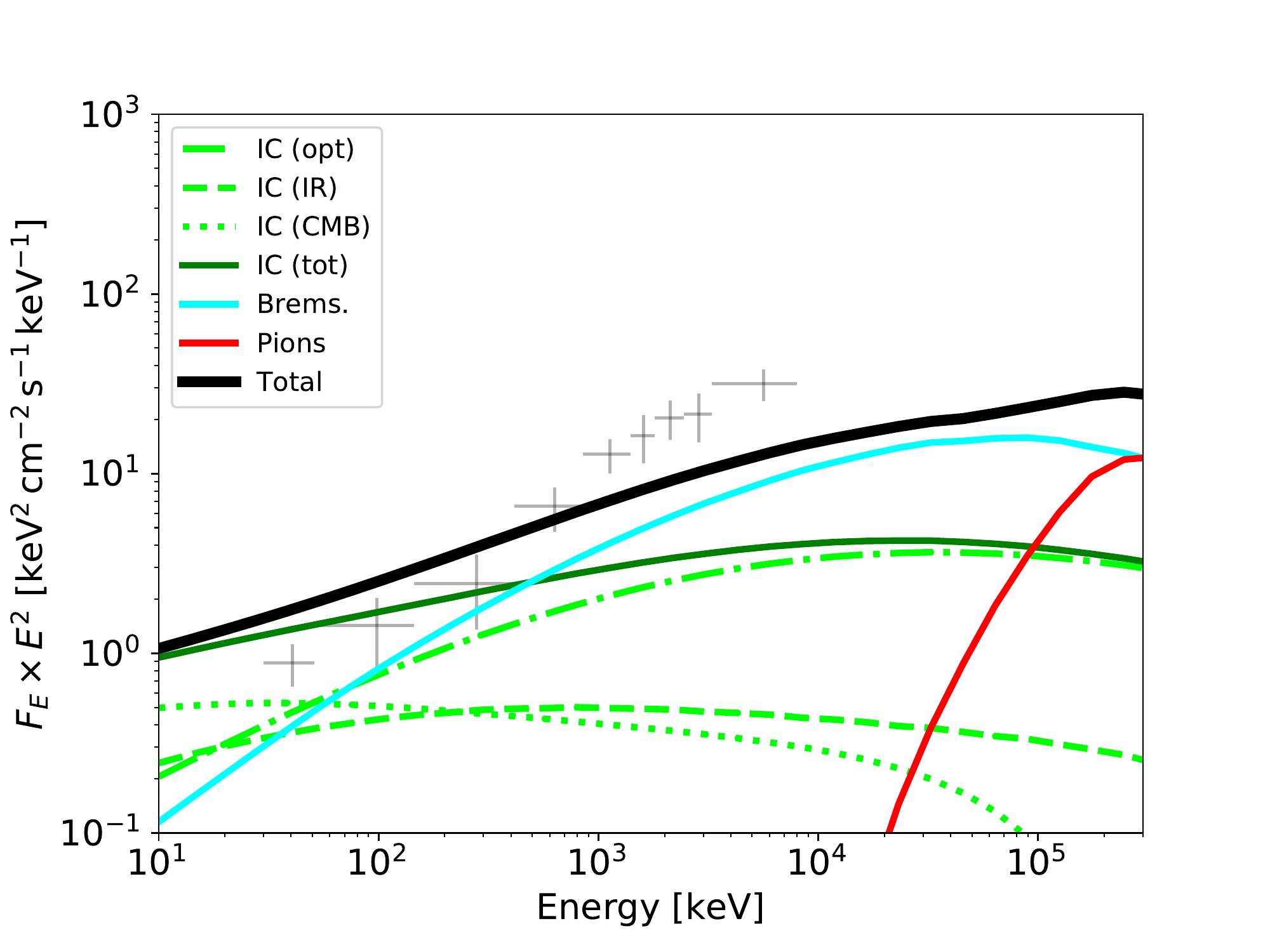}
	\includegraphics[width=0.49\textwidth]{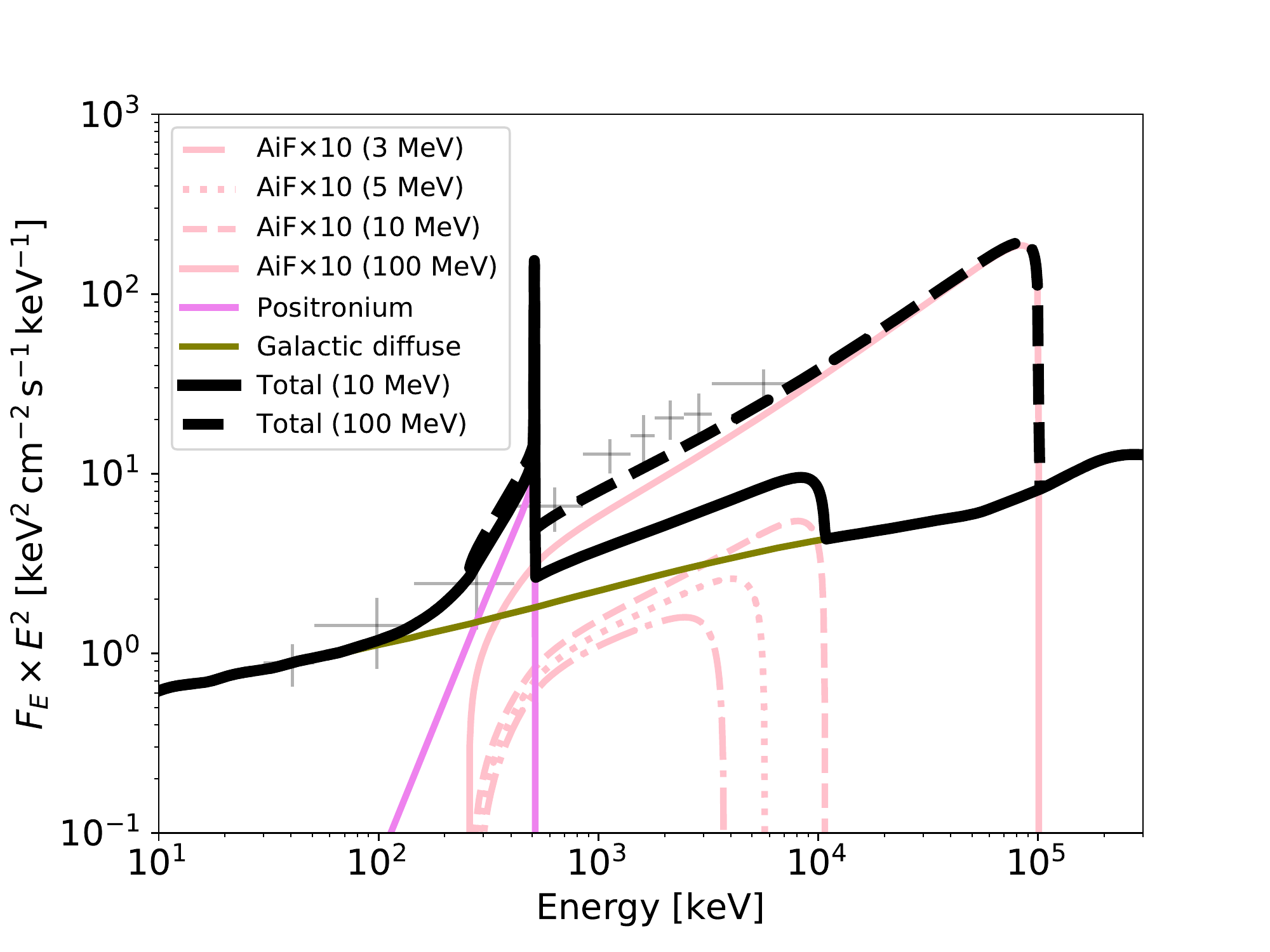}\\
	\includegraphics[width=0.49\textwidth]{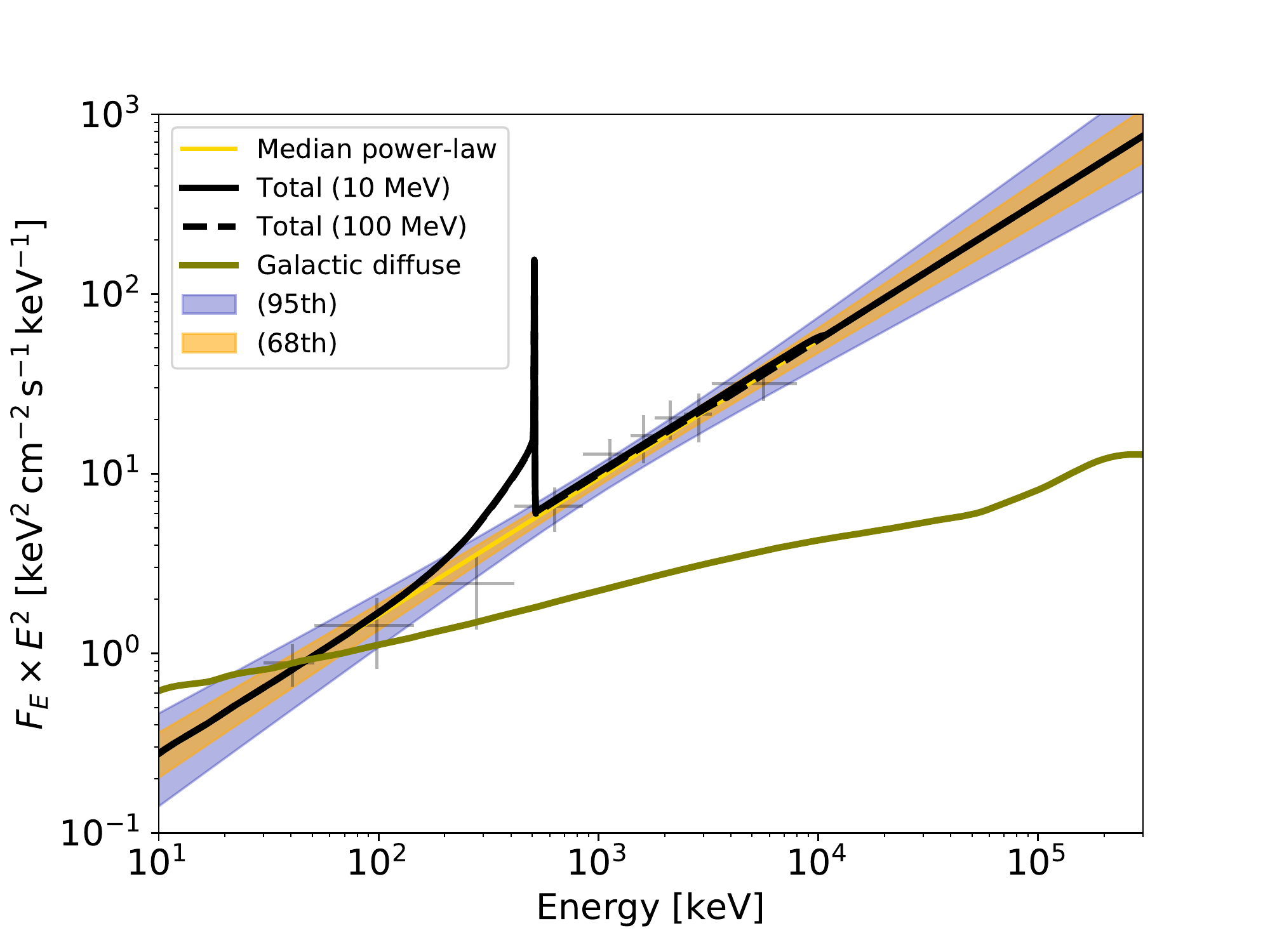}
	\includegraphics[width=0.49\textwidth]{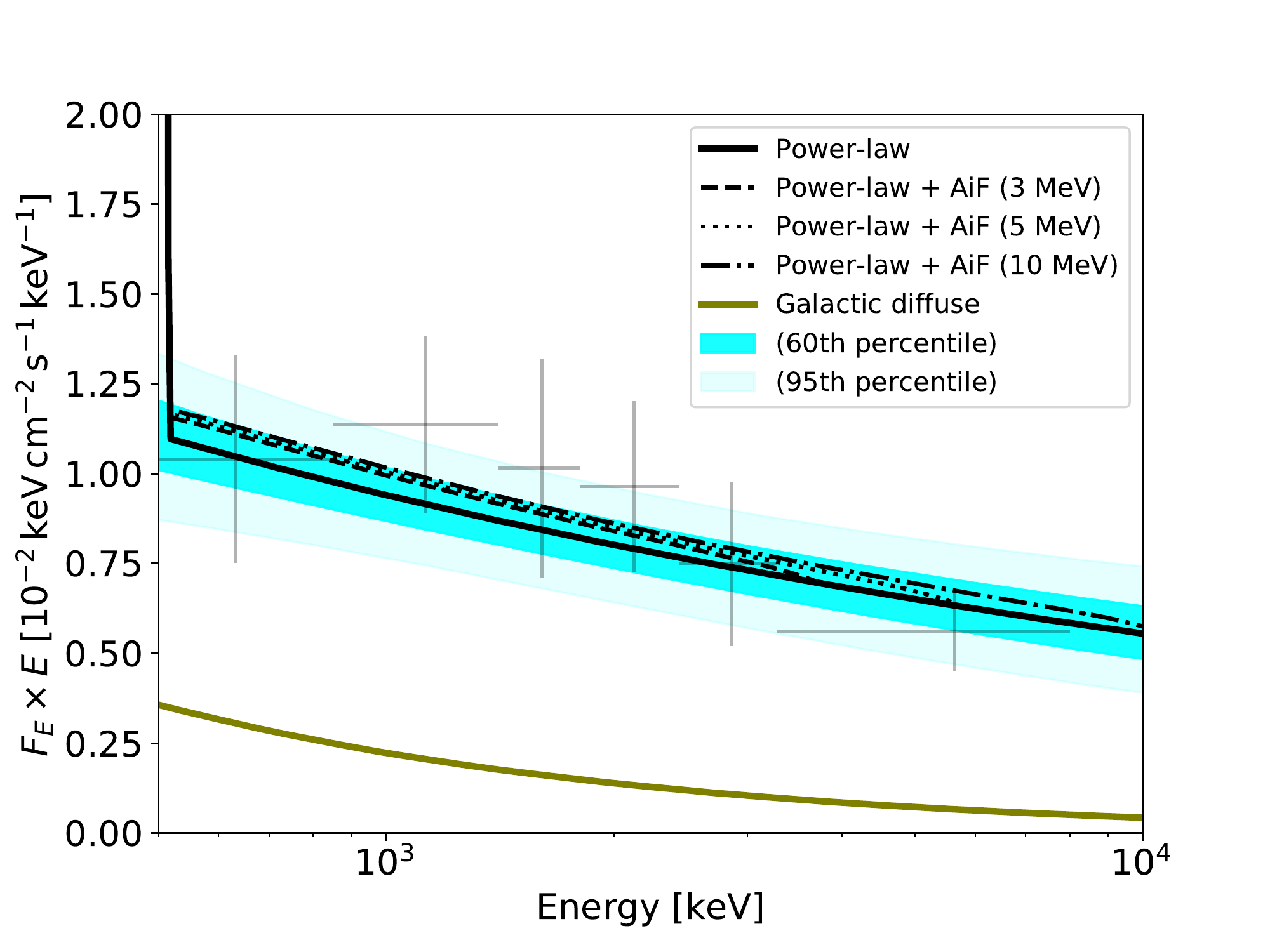}
	\caption{Diffuse MeV continuum spectrum possibilities of the Milky Way in the range $|l| \leq 47.5^{\circ}$, $|b| \leq 47.5^{\circ}$. The gray data points indicate the measured Milky Way spectrum between 0.03 and 8\,MeV with INTEGRAL/SPI \citep{Siegert2022_MWdiffuse,Berteaud2022_SPI_PBH}, excluding the $\gamma$-ray lines at 511\,keV and 1809\,keV, as well as the ortho-Ps continuum. \emph{\textbf{Top left:}} Generic GALPROP model components can roughly explain the entire spectrum with Inverse Compton (IC) emission from the CMB, infrared (IR) light, and optical (opt, scaled up by a factor of 2, suggested by \citet{Bouchet2011_diffuseCR}, for example) light, and a bremsstrahlung component (scaled up by a factor of 3 to roughly match the data). \emph{\textbf{Top right:}} The Galactic diffuse baseline model (no scalings) allows for additional components, such as \pos{}-annihilation in flight. A ten times stronger than measured flux of \pos{}-annihilation allows annihilation in flight up to injection energies above 100\,MeV. \emph{\textbf{Bottom left:}} A power-law description (shaded bands for $1$ and $2\sigma$ uncertainties) would explain the data completely, but would disagree with a Galactic diffuse baseline model. \emph{\textbf{Bottom right:}} Zoom in to the fit from the bottom left plot, in the style of \citet{Beacom2006_511}, but extended to the entire region of interest. Assuming the fitted power-law explains the entire emission, within uncertainties, the annihilation in flight injection energy is automatically constrained to below 10\,MeV. Work in progress / preliminary results by Siegert et al. (2023, in prep.).}
	\label{fig:IC_AiF_specs}
\end{figure*}

About 1): If it is assumed that \poss{} are injected mono-energetically with initial energy $E_i$ (which is never the case, except for maybe DM), in an ISM of number density $n_{\rm H}$ and temperature $\lesssim 10^5$\,K, with Lorentz-factors $\gamma \lesssim 100$, the dominant cooling process is ionisation and excitation of ambient gas.
If only this process is considered, then the survival probability, $P$, i.e. not annihilating in flight, is always greater than 80\,\%, rising with decreasing kinetic energy.
The in-flight annihilation flux, $F_{\rm IA}$, is calculated as \citep{Beacom2006_511}
\begin{equation}
	F_{\rm IA} = F_{511} \frac{1}{1 - \frac{3}{4}f_{\rm Ps}}\frac{1-P}{P} := r_{\rm IA} F_{511}\mrm{,}
	\label{eq:inflight_flux}
\end{equation}
where $f_{\rm Ps}$ is the (measured) Ps fraction from the ratio of the 511\,keV line and the ortho-Ps continuum, $0 \leq r_{32} \leq 4.5$ \citep{Ore1949_511}.
For a canonical value of $f_{\rm Ps}$ close to 1 as suggested by measurements, the total in-flight annihilation flux is at most the 511\,keV line flux, but distributed over a wide continuum (see Fig.\,\ref{fig:IC_AiF_specs}).
The smaller the injection energy, the smaller the total in-flight annihilation flux, and the less pronounced is the high-energy bump corresponding to the injection energy (minus $m_e/2$).
However, this is not the only energy loss possible in the ISM:
If the kinetic energy of the \poss{} is smaller than the plasma temperature of, say $k_BT \lesssim 10$\,keV, temperature-dependent losses set in, and can be orders of magnitude larger than the ionisation losses.
Clearly, the ionisation fraction would also be close to one, replacing ionisation losses with Coulomb losses.
Such high temperatures may be found in hot gas with lower densities such as in superbubbles or the Galactic halo, lowering the survival probability and the resulting normalisation of the in-flight annihilation flux.

In addition to the potentially increased flux for annihilation in flight, the premise of `all \poss{} are annihilating' may not hold in the light of INTEGRAL/SPI observations:
As stated above, coded aperture masks can only hardly observe isotropic emission, such as would be plausible for a halo component.
Hydrodynamics simulations suggest that radioactive isotopes, and therefore the starting points of \pos{}-propagation, can reach very high distances above the Galactic plane \citep{Krause2021_26Alchimneys}.
If \poss{} are emitted at heights of 2\,kpc or higher, they may propagate freely in the halo medium and eventually annihilate on the border to the IGM.
Unless the \poss{} are guided back to Galaxy \citep[e.g.,][]{Prantzos2006_511}, they have no preferential target to annihilate with.
This could then result in an isotropic component but which would be invisible for current $\gamma$-ray telescopes.
As suggested in Sec.\,\ref{sec:sources}, the production rate in the Milky Way might be a factor of five or more larger than what is \emph{seen} to annihilate.
If such an additional component exists, the \pos{}-annihilation in flight spectrum might in fact be larger by about this factor (integrated over the whole sky, but with a possibly different morphology).

About 2): It has been suggested that the injection energy of \poss{} into the ISM must not be larger than 3--7\,MeV, based on model comparisons to already-extracted flux data points without taking into account the energy dispersion of the used $\gamma$-ray instruments \citep[e.g.,][]{Beacom2006_511,Sizun2006_511}.
This claim originates from the fact that it is assumed that the diffuse $\gamma$-ray continuum emission in the Milky Way is understood, and only taking into account a region with a radius of $5^{\circ}$ around the Galactic centre.
Similar to the case of high-energy cosmic-ray \poss{} (see Sec.\,\ref{sec:high-energy-propagation}), the former is not the case.

In particular, it was assumed that the MeV spectrum can be described by a power-law, which is certainly the case, given the uncertainties of the data points (see Fig.\,\ref{fig:IC_AiF_specs}).
However, searching for an additional component \emph{above} this power-law without allowing to change the power-law parameters in a spectral fit, will never find an additional component, because the power-law \emph{already} describes the data well.
This means an annihilation in-flight spectrum has to be as small as to not `overshoot' the power-law model, which results in an inevitable upper bound on the injection energy of 3--7\,MeV, because the total flux is bound to the 511\,keV line flux.

Now, if an astrophysical model is assumed for the diffuse $\gamma$-ray continuum, for example calculated with GALPROP \citep{Strong2007_GALPROP}, the model is parametrised with the cosmic-ray \ele{} spectrum, i.e. its spectral shape and amplitude, with the interstellar radiation field, with the particle density (for bremsstrahlung and pion-production), and with the magnetic field, among other parameters.
Such a parametrisation allows for a much more flexible model, and, given the uncertainties of the measurements in the MeV range as well as from direct cosmic-ray measurements and high-energy $\gamma$-rays, results in a less-determined spectrum than by just fitting a power-law (see Fig.\,\ref{fig:IC_AiF_specs}).
The reason why the \pos{} injection energy appears to be bound to several MeV is because a model is assumed that already fully explains the data.

In Fig.\,\ref{fig:IC_AiF_specs}, I show four cases that are compatible with the MeV data and still allowed by other constraints from cosmic rays and GeV $\gamma$-rays.
It is clear that from the MeV $\gamma$-ray spectrum alone, the annihilation in flight spectrum and therefore injection energies, cannot be determined -- at least with current instrumentation.
A precise statement requires a broader spectral range, must also take into account the measured cosmic-ray \ele{} spectrum at GeV energies, and must allow for different cosmic-ray populations in the Galaxy.

\section{Future avenues}\label{sec:future}
The Positron Puzzle appears to require a smoking-gun-evidence measurement to proceed.
But solving the puzzle does not only rely on better measurements with future instrument -- of course the latter will tremendously help --, it also relies on understanding the MeV sky itself better as well as more solid data analyses.

\subsection{Foreground emission}\label{sec:foreground}
While the 511\,keV line is the strongest astrophysical $\gamma$-ray line known, the will to take `pictures' of the MeV sky unavoidably results in ambiguities, image reconstruction artefacts, and finally doubt about the results.
But many of the low-significance features in MeV images might in fact not be artefacts but due to a time variable foreground emission.
As described in Sec.\,\ref{sec:astro_sources}, the cosmic-ray bombardment of small Solar System bodies leads to nuclear excitation and pion production, followed by de-excitation via $\gamma$-ray emission and \pos{}-annihilation on top of a bremsstrahlung spectrum.
It is known from $\gamma$-ray measurements of the asteroid 433 Eros \citep{Peplowski2016_Eros433}, that there is indeed a 511\,keV line, plus de-excitation lines that inform about the asteroid's composition.
From the entire population of asteroids in the main belt between Mars and Jupiter (torus at 2.0--3.5\,AU), the Kuiper belt beyond Neptune (torus at 30--50\,AU), the Jovian trojans (accumulation at Lagrange points of the Jupiter-Sun-system; orbital time 11\,yr), the plausible Neptunian trojans (orbital time 165\,yr), and the suspected but not proven Oort-cloud (spherical accumulation between $10^3$--$10^5$\,AU) would all contribute to a time variable foreground emission, predominantly in the ecliptic \citep[see also][for estimates]{Moskalenko2008_GRalbedoSS}.
The Oort-cloud, if real, would even lead to an isotropic 511\,keV, and in general broadband MeV, emission.
For coded aperture masks, this emission would be invisible, but would be detected by Compton telescopes, such as the Compton Spectrometer and Imager \citep[COSI,][planned for launch in 2027]{Tomsick2019_COSI}, if the flux is large enough.
The ecliptic has an intriguing overlap with the Galaxy, almost exactly at the Galactic bulge, so that the bulge emission at 511\,keV with a flux of $10^{-3}\,\mrm{ph\,cm^{-2}\,s^{-1}}$ could actually be expected to change as a function of planetary time scales by 1--5\,\%.
Such a measurement is on the verge of possibilities with current instrumentation.
Since the ecliptic is rarely observed with INTEGRAL/SPI, the possibility to detect the foreground emission directly requires a time variable diffuse emission model.
If emission from apparently moving sources is analysed time independent, their resulting picture, or map, may include high-flux regions which can hardly be explained by astrophysical sources outside the Solar System.

The 511\,keV line sensitivity we would need to identify the time-averaged components we probably already have with INTEGRAL/SPI, about $8 \times 10^{-5}\,\mrm{ph\,cm^{-2}\,s^{-1}}$ ($3\sigma$ point source sensitivity for 1\,Ms).
But the current exposure of SPI, being mostly concentrated along the Galactic plane and bulge, makes it difficult to identify any high-latitude component.
Since the individual components would be large on the sky, possibly with several tens of degrees in diameter, the line sensitivity to disentangle these foreground components should be below $10^{-5}\,\mrm{ph\,cm^{-2}\,s^{-1}}$.
The future COSI-SMEX mission \citep{Tomsick2019_COSI} will shed light on the Solar System foreground.

\begin{figure*}[t]
	\centering
	\includegraphics[width=0.49\textwidth]{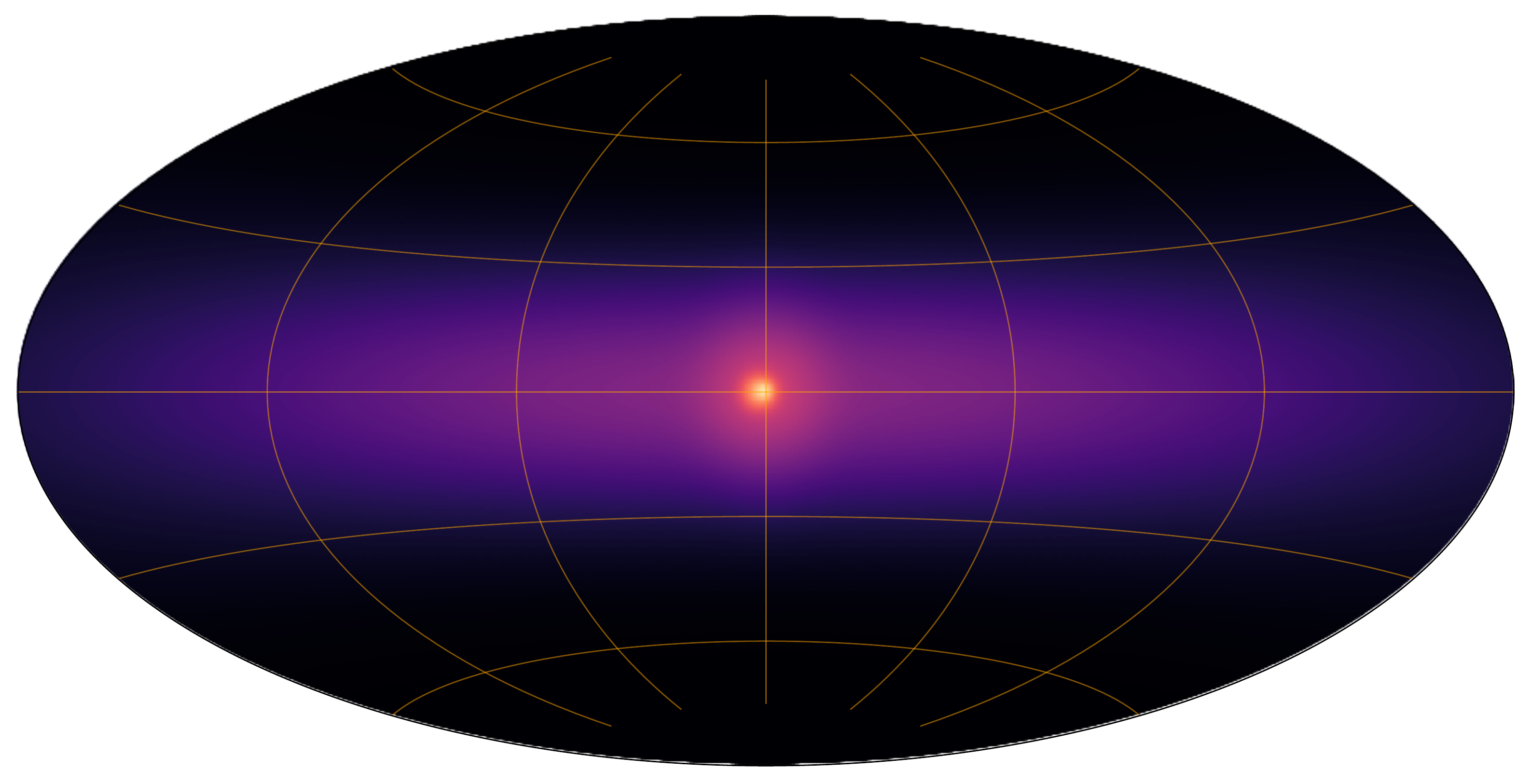}	
	\includegraphics[width=0.49\textwidth]{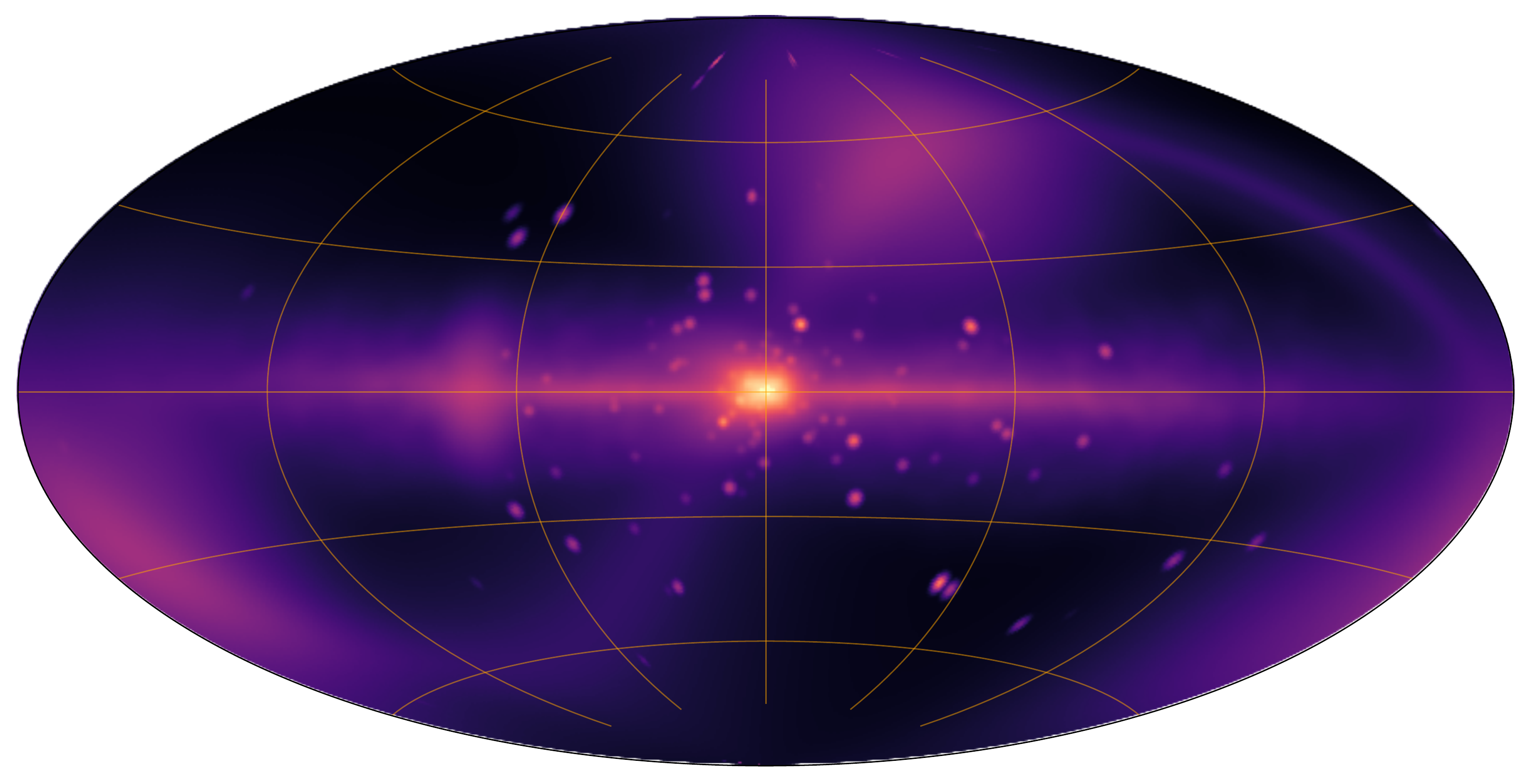}
	\caption{Positron annihilation in the Galaxy. \emph{\textbf{Left:}} Four-component model from \citet{Siegert2016_511} that describes the INTEGRAL/SPI data completely. \emph{\textbf{Right:}} Astrophysically motivated model of annihilation emission, including old stars (boxy bulge, nuclear stellar cluster and disk, infrared disk; intermittently flaring stars), globular clusters (point-like), spiral arm structure (HI annihilation regions), dominant massive star regions (e.g. Cygnus OB2, Orion), plus foreground emission (main belt asteroids and Jovian trojans) along the ecliptic. Dark matter halo and isotropic components are not shown. The sensitivity (lower flux limit) of the right image is about a factor of 20 better than in the left image.}
	\label{fig:old_vs_new_image}
\end{figure*}

\subsection{Isotropic Emission}\label{sec:isotropic}
In addition to the Oort cloud, there may be even more isotropic components, especially for the 511\,keV line.
From the Nested Leaky-Box-Model of \citet{Cowsik2014_CRpropagation}, it is clear that a large portion of \poss{} may leave the Galaxy.
But also in the Galactic halo and the IGM, \poss{} would eventually find their \eles{} to annihilate with.
Depending on the temperature where \poss{} recombine with \eles{}, the 511\,keV line would be broadened accordingly.
Given the direct measurements, and accounting for all uncertainties, even within $1\sigma$, more than five times the \poss{} \emph{seen to annihilate} could be produced in the Milky Way.
Within $2\sigma$, up to a factor of ten is possible.
Only with a Compton telescope or vastly improved coded-mask apertures, or in general MeV telescopes, could such an isotropic component be detected from the Milky Way.

A similar isotropic emission component at 511\,keV might be expected from DM, as the line-of-sight integration into all directions always leads to an offset, which would result in an isotropic part.
From the CGB, such a component would then also include a red-shift dependence, so that both, a strong 511\,keV line, and a continuum-like component due to integration across red-shifts would be expected \citep{Iguaz2021_CGBPBH}.

Finally, also the pair-absorption of very- and ultra-high-energy photons will create \ele{}-\pos{}-pairs in the Galaxy and the IGM which will eventually annihilate and plausibly produce an isotropic 511\,keV line.
The mean free path of PeV photons, for example, is less than 10\,kpc so that any `pevatron' inside the Galaxy will also create \poss{} through pair production with the CMB and the interstellar radiation field.
For extragalactic sources, this also applies, and the shear number of extragalactic $>$TeV sources will lead to the inevitable pair production in the IGM.

The possibility of an isotropic 511\,keV line is an going research topic and there are hardly any flux expectations or predictions.
As mentioned above, coded aperture masks can hardly identify isotropic emission unless specific satellite orbit manoeuvres are performed.
INTEGRAL Earth occultation observations allowed us to measure the cosmic X-ray background up to $\sim 200$\,keV, but did not consider the 511\,keV line \citep[e.g.][]{Churazov2007_CXB_INTEGRAL,Tuerler2010_CXB_INTEGRAL}.
A current all-sky sensitivity of SPI of $10^{-2}\,\mrm{ph\,cm^{-2}\,s^{-1}}$ is estimated from the previous results.
A ten-fold sensitivity improvement in the 511\,keV line might already show if there are \poss{} annihilating in the halo which current instruments are blind for.

\subsection{Imaging might not be enough}\label{sec:powerspectrum}
With new MeV telescopes, i.e. improved angular resolution and better sensitivity, new images of the 511\,keV sky will be created.
But even with the identification of 511\,keV point sources, there may be open questions, as to why the morphology is what it is.
The comparison of tracer maps, and ranking them according to relative likelihoods gives an idea of what the sources or sinks are, but the physics cannot be probed on this basis.

\citet{Pleintinger2019_26Al} tried to disentangle the overall morphology of the \nuc{Al}{26} line at 1809\,keV by slicing the data into longitude-bins and measuring the scale height by a maximum likelihood fit.
The regions were chosen independent from each other, so that a statistic about scale heights could be generated.
From Hydrodynamics simulations, for example, or population synthesis calculations, a similar statistic can then be calculated which includes all the astrophysical input.
From this treatment it was found that indeed, chimney-like structures could exist in the Galaxy, which was not possible before.

A more advanced approach would be a decomposition similar to the CMB power spectrum.
If the raw data, or in a weaker version the reconstructed images, can be analysed through spherical harmonics, a meaningful multipole series expansion up to $l \approx 100$ would be possible with future telescopes.
In this way, the physical interpretation does not rely dramatically on the background model, and inferences from costly hydrodynamics simulations are made possible.
Especially in the case of the 511\,keV emission with its many source distributions, the full (unknown) propagation, and final annihilation sites, will benefit from structured, high-resolution, simulations to understand the full impact of the puzzle.

Structural comparisons will be better the better the sensitivity of the instrument and the more homogeneous the exposure.
Size and angular scales can be more solidly compared if the sensitivity is improved by another decade with respect to INTEGRAL/SPI, for example.
The COSI-SMEX mission will have a 511\,keV line point source sensitivity of less than $10^{-5}\,\mrm{ph\,cm^{-2}\,s^{-1}}$ \emph{for each position in the sky} after its two-year nominal mission, thanks to its large field of view of $1\,\pi\,\mrm{sr}$ \citep{Tomsick2019_COSI}.

\subsection{What gives?}\label{sec:other_problems}
With new measurements from a new telescope, further analyses towards understanding the MeV foreground, and data analysis tools that focus on the astrophysical parameters, the Positron Puzzle may solve more than just the origin of \poss{} in galaxies.
With current instrumentation, it is already now possible to develop astrophysical models that are bound by the total emission in the Galaxy.
The 511\,keV line, and associated also the ortho-Ps continuum as well as the in-flight annihilation component, can be used  to study other unsolved problems in astrophysics.
I will outline two examples here that have, at first glance, nothing in common, but are indeed connected through \pos{}-annihilation.

\paragraph{Dark matter:}
As described above, DM in the form of beyond-standard-model particles or PBHs will eventually lead to non- or highly-relativistic \poss{}, or a mixture in between.
These \eles{} and \poss{} will be responsible for additional spectral components from GHz to GeV.
As the PBHs evaporate or DM particles co-annihilate, the secondary particles will decay and result in a prompt emission spectrum around the scale of the PBH or DM mass, typically peaking at about 10\,\% of the particles' masses.
The secondary high-energy \eles{} and \poss{} created propagate diffusively and experience losses from IC scattering off of CMB and infrared/optical photons \citep[e.g.,][]{Saxena2011_ICdarkmatter,Cirelli2009_IC_darkmatter}.
This would create a characteristic IC bump that should be consistent throughout all astrophysical observation possibilities (Milky Way, dwarf galaxies, isotropic with redshift integration).
For GeV pairs, this would result in peaks in the MeV range (see Fig.\,\ref{fig:IC_AiF_specs}).
In addition will the pairs experience synchrotron losses, so that a characteristic synchrotron spectrum would be visible, also with a peak that depends on the pairs' energies and hence the initial DM particle mass.
When the pairs are then cooled and the IC, synchrotron and bremsstrahlung losses are sub-dominant, Coulomb losses set in, and \eles{} and \poss{} require on the order of 0.1--10\,Myr to cool down to thermal temperatures (see Sec.\,\ref{sec:low-energy-propagation}).
During this entire time, the \poss{} may also annihilate in flight, leading to another characteristic feature that depends on the initial \pos{} spectrum.
Then, after thermalisation, \poss{} eventually annihilate via the formation of Ps, showing the strong 511\,keV line and the ortho-Ps component.

Building a full model of the secondary and tertiary particles from DM annihilation (or PBH evaporation) in the Milky Way, and adequately comparing this to the available data from radio to $\gamma$-rays is difficult as the astrophysical foreground has to be understand.
But especially the IC MeV continuum and 511\,keV line are orthogonal to most other wavelengths (see Fig.\,\ref{fig:line_map_wavelengths}) in which a `prompt' DM annihilation may be visible.
This makes them an invaluable messenger in the indirect search for DM.
The cosmic-ray \poss{} and \eles{}, for example measured with AMS-02, should also be used to constrain the DM searches \emph{in addition} to photon observations -- not independently.

\begin{figure*}[t]
	\centering
	\includegraphics[width=1\textwidth]{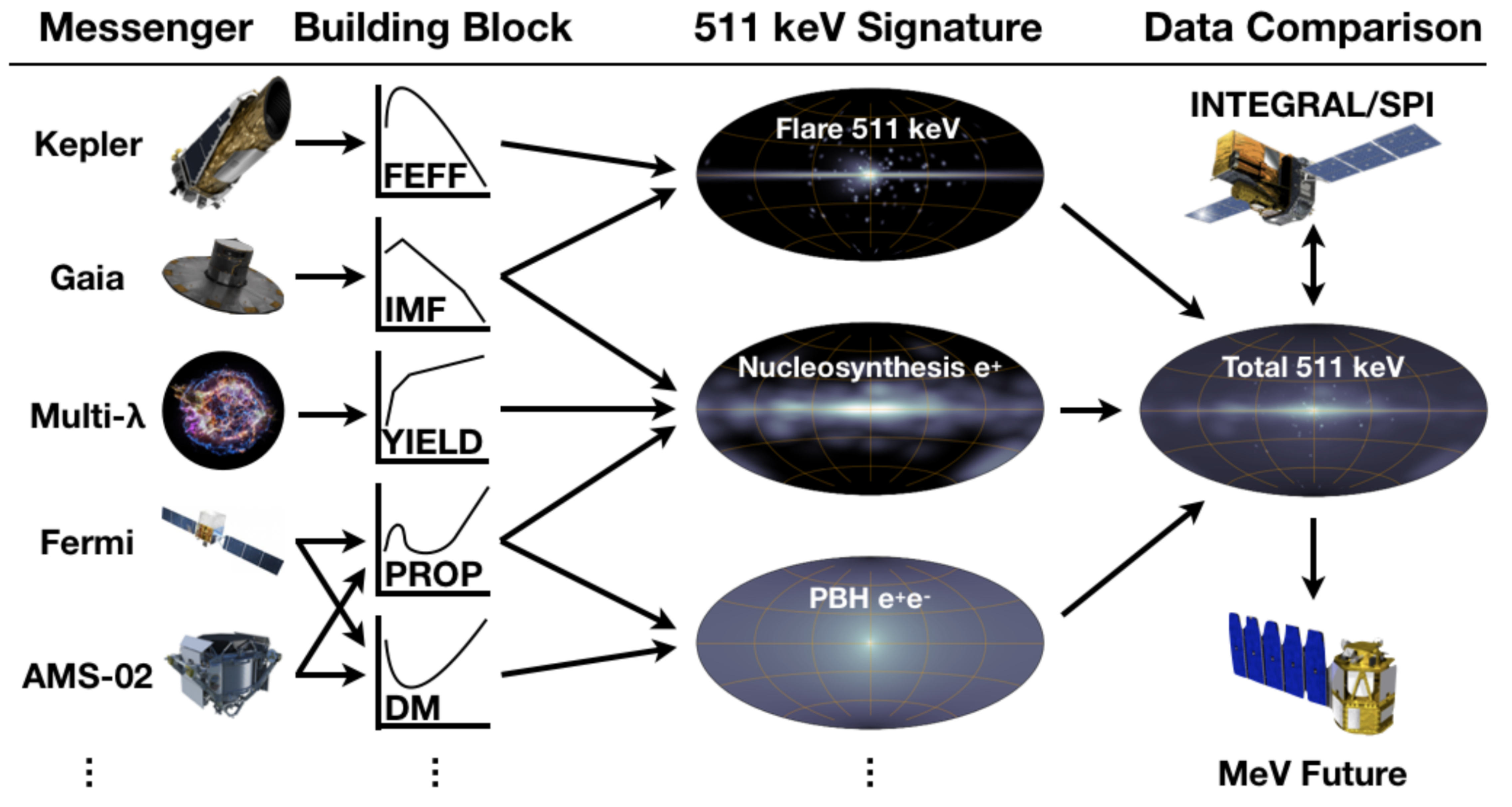}
	\caption{Model and observational components required to solve the Positron Puzzle. From left to right, the messengers (multi-wavelength and cosmic-ray measurements) inform the building blocks of a intertwined (hierarchical) model, that provides the flare-energy-frequency function (FEFF), the IMF, the nucleosynthetic yields (of $\beta^+$-decayers), the propagation parameters, and DM limits, among others. The building blocks are used to create \pos{}-source-specific images at 511\,keV and all other components (ortho-Ps, annihilation in flight, IC, etc.), which can then be fitted to the raw data of current (INTEGRAL/SPI) and future (e.g., COSI) instruments sensitive to \pos{}-annihilation.}
	\label{fig:hierarchical_model}
\end{figure*}

\paragraph{Solar coronal heating problem:}
Assuming instead that the 511\,keV emission that is coincident with the positions of stars (i.e. a considerable fraction, see Fig.\,\ref{fig:likelihood_comparisons}) comes from stellar flares, one can build a model to determine the lower and upper cut-off energy of the flare-frequency-energy distribution.
Determining the lower cut-off energy from the entire population of stars may inform the solar coronal heating problem as it is believed that micro- and nano-flares are responsible for the high temperatures of the Sun's corona.
Instead of observing a single star in 511\,keV, which is only possible for the Sun, the Galactic 511\,keV emission provides a means to observe \emph{all} stars at the same time.
For such an endeavour, one needs to build a hierarchical model that eventually predicts the total flux at 511\,keV (and ortho-Ps).
Again, for the entire Milky Way, such a model will be difficult, but hotspots, such as expected from globular clusters, would result in characteristic ratios among the 157 globular clusters of the Milky Way \citep{Harris2010_GCcatalogue}.
Such a ratio will be very rigid so that different models, which may also be tested in globular clusters \citep{Bartels2018_binaries511}, could easily be distinguished.

In particular, for each of the $N=j$ globular clusters of the Milky Way, an initial mass function (IMF) with shape $\Gamma_j$ and total number of stars $N_{j*}$ determines the masses $M_*$ of stars in each object.
Each star with mass $M_*$ has its own flare-frequency-energy distribution that determines the average number of \poss{} in each flare.
The flare distribution depends on three parameters, the low-energy cut-off, $E_{\rm lo}$, high-energy cut-off $E_{\rm hi}$, and a steepness parameter $\alpha$, where $E_{\rm lo}$ and $E_{\rm hi}$ could also depend on the stellar mass.
The star's mass also determines the density profile, $n_{*}$, of the stellar atmosphere, from which finally the number of annihilation photons (flux) can be calculated.
Given the distances to all globular clusters in the Milky Way, a flux pattern of 511\,keV point sources would be result (see Fig.\,\ref{fig:old_vs_new_image}, right).
The only fitted parameters of such a model would be $E_{\rm lo}$ and $E_{\rm hi}$ since other parameters would be determined from independent measurements.
These cut-offs serve as normalisation parameters for the final 511\,keV flux because if $E_{\rm lo}$ is too small or $E_{\rm hi}$ too large, too many stars would flare too frequently, so that the globular clusters would appear brighter than what is actually measured.

Based on a correlation with the GeV flux \citep{Bartels2018_binaries511}, one can estimate to an order of magnitude how bright several objects would be in the 511\,keV line.
There could be 19 globular cluster with a flux of $10^{-6}\,\mrm{ph\,cm^{-2}\,s^{-1}}$ or more, with Omega Centauri, 47 Tucanae, Terzan 5, and NGC 6397 as potentially the brightest ones with fluxes of $(1$--$2) \times 10^{-5}\,\mrm{ph\,cm^{-2}\,s^{-1}}$.
COSI would easily detect these objects within its two year mission \citep{Tomsick2019_COSI}.
The total 511\,keV flux of all globular cluster combined would be around $2 \times 10^{-4}\,\mrm{ph\,cm^{-2}\,s^{-1}}$.
Only a few objects would be seen individually, but since they follow a certain hierarchy given their GeV fluxes, they can be treated as one population and discovered as one entity.

For a stellar flare scenario \citep{Bisnovatyi-Kogan2017_511}, i.e. roughly scaling with stellar mass, the expected fluxes would be mostly below $10^{-6}\,\mrm{ph\,cm^{-2}\,s^{-1}}$ for individual objects.
Again Omega Centauri and 47 Tucanae would be the brightest, but not individually detectable, even with COSI.
The total flux of all 157 objects would be on the edge of the detection limit for future instruments in the case of a 511\,keV line purely from stellar flares.

\section{Putting it all together}\label{sec:conclusion}
It is clear that both exemplary models from above, DM and stellar flares, may in fact contribute to the entire Galactic signal.
Extrapolating the models from individual objects to the entire Milky Way signal will turn out to be more complex as they might be interdependent -- but are bound by the morphology and total flux measured.
Such models, if \pos{}-transport is properly taken into account, can actually be \emph{fitted} to data instead of just building the advanced models.
The resulting 3D emissivities need to be line-of-sight integrated which leads to 2D-images, parametrised by only a few parameters, similar to heuristic models but with physical meaning.
The final all-sky images (i.e. the Milky Way, plus \emph{all} other contributions from dwarf galaxies, globular clusters, nearby sources, isotropic emission), and spectral differences (not just the 511\,keV line but annihilation in hot gas, Ps fraction, annihilation in flight, IC components, etc.) can be directly compared to raw data of instruments.

In order to build a full understanding of the Positron Puzzle, it cannot be solved by arguing about individual sources -- they must all be described together.
This means that for each \pos{}-source population, a realistic \pos{}-production mechanism, escape, and injection to the ISM needs to be built.
But all these source populations and injected \poss{} are required to experience the same propagation mechanism through which the final annihilation image is tied together.
If the \poss{} do not escape, a fraction of the Galactic 511\,keV emission will come from in-situ annihilation, for example from stellar flares and SN remnants.

The two important messages to understand here are:
1) Occam's razor must not be misinterpreted in the case of finding \emph{the} source responsible for annihilating \poss{} in the Milky Way.
It is clear from a multitude of observational evidence that there must be more than one source that contributes to the entire measured signal.
2) Only by considering more data than just the 511\,keV line will lead to reasonable progress in solving the Positron Puzzle (Fig.\,\ref{fig:hierarchical_model}).
In turn, by having this unique astrophysical messenger, we will gain insights to other unsolved astrophysical problems.

\backmatter

\bmhead{Supplementary information}
Not applicable

%If your article has accompanying supplementary file/s please state so here. 

%Authors reporting data from electrophoretic gels and blots should supply the full unprocessed scans for key as part of their Supplementary information. This may be requested by the editorial team/s if it is missing.

%Please refer to Journal-level guidance for any specific requirements.

\bmhead{Acknowledgments}
I thank Martin G. H. Krause, Karl Mannheim, Roland Diehl, Jochen Greiner, John Tomsick, and Steve Boggs for support throughout the years and for fruitful discussions.

\section*{Declarations}
Not applicable.

%Some journals require declarations to be submitted in a standardised format. Please check the Instructions for Authors of the journal to which you are submitting to see if you need to complete this section. If yes, your manuscript must contain the following sections under the heading `Declarations':

%\begin{itemize}
%\item Conflict of interest/Competing interests (check journal-specific guidelines for which heading to use)
%\item Ethics approval 
%\item Consent for publication
%\item Availability of data and materials
%\item Authors' contributions
%\end{itemize}

%\noindent
%If any of the sections are not relevant to your manuscript, please include the heading and write `Not applicable' for that section. 

\bigskip
\begin{flushleft}%
Editorial Policies for:

\bigskip\noindent
Springer journals and proceedings: \url{https://www.springer.com/gp/editorial-policies}

\bigskip\noindent
Nature Portfolio journals: \url{https://www.nature.com/nature-research/editorial-policies}

\bigskip\noindent
\textit{Scientific Reports}: \url{https://www.nature.com/srep/journal-policies/editorial-policies}

\bigskip\noindent
BMC journals: \url{https://www.biomedcentral.com/getpublished/editorial-policies}
\end{flushleft}

%\begin{appendices}

%\section{Section title of first appendix}\label{secA1}

%An appendix contains supplementary information that is not an essential part of the text itself but which may be helpful in providing a more comprehensive understanding of the research problem or it is information that is too cumbersome to be included in the body of the paper.




%\end{appendices}

\bibliographystyle{sn-basic}
\bibliography{thomas.bib}% common bib file


\end{document}